\documentstyle[12pt,epsf]{article}

\begin{document}

\baselineskip=18.8pt plus 0.2pt minus 0.1pt

\makeatletter

\renewcommand{\thefootnote}{\fnsymbol{footnote}}
\newcommand{\beq}{\begin{equation}}
\newcommand{\eeq}{\end{equation}}
\newcommand{\bea}{\begin{eqnarray}}
\newcommand{\eea}{\end{eqnarray}}
\newcommand{\nn}{\nonumber \\}
\newcommand{\hs}[1]{\hspace{#1}}
\newcommand{\vs}[1]{\vspace{#1}}
\newcommand{\Half}{\frac{1}{2}}
\newcommand{\p}{\partial}
\newcommand{\ol}{\overline}
\newcommand{\wt}[1]{\widetilde{#1}}
\newcommand{\ap}{\alpha'}
\newcommand{\bra}[1]{\left\langle #1 \right\vert }
\newcommand{\ket}[1]{\left\vert #1 \right\rangle }
\newcommand{\vev}[1]{\left\langle #1 \right\rangle }
\newcommand{\vac}{\ket{0}}

\newcommand{\ul}[1]{\underline{#1}}

\makeatother

\begin{titlepage}
\title{
\hfill\parbox{6cm}
{\normalsize CERN-PH-TH/2004-118\\{\tt hep-th/0406242}}\\
\vspace{1cm}
On-shell Gauge Invariants and Field Strengths in 
Open Superstring Field Theory
}
\author{Yoji Michishita
\thanks{
{\tt Yoji.Michishita@cern.ch}
}
\\[7pt]
{\it Theory Division, CERN,
CH-1211, Geneva 23, Switzerland}}

\date{\normalsize June, 2004}
\maketitle
\thispagestyle{empty}

\begin{abstract}
\normalsize
We study gauge invariant quantities in the open superstring field theory
proposed by Berkovits, extending the precedent discussion
in bosonic string field theory. Our gauge invariants are ``on-shell''.
As its applications, we define quantities which are expected to be
related to the U(1) field strength --- a RR coupling and a ``component'' 
of the string field equation of motion, and consider their naive extensions
to off-shell.
Order by order calculations show that the field strength extracted from the
RR coupling is not gauge invariant, while from the component of the equation 
of motion we obtain an off-shell field strength which is gauge invariant
under full gauge transformation if on-shell,
and under linearized gauge transformation even off-shell.
\end{abstract}



\end{titlepage}

\section{Introduction}

Open string field theory has been developed in the last several years, 
mainly in the study of tachyon condensation. However it is not yet a 
tool as useful as first quantized formalism when we consider other phenomena.
Further development is necessary
to make it as useful as first quantized formalism.
Especially, in this theory it is very difficult to extract physical 
quantities, because of its computational complexity and
nonlocality. It is helpful to know how to construct physically 
meaningful quantities. 

String field theory is a kind of gauge theory, and gauge invariant quantities
in gauge theories have some physical meanings. Therefore we want
to construct gauge invariant quantities in string field theory. 
Studies along this direction in bosonic string field theory
have been advanced in \cite{z,grsz,hi}. These works have shown that,
roughly speaking, couplings of one closed string mode and one open string 
field are gauge invariant. In this paper we study these gauge invariants
in open superstring field theory proposed by Berkovits \cite{b1}.
This gauge invariant is ``on-shell'' because of the on-shell property of 
the closed string mode. 

The simplest gauge invariant quantity in U(1) gauge theory 
is the field strength of the gauge field. Since in lowest order approximation 
U(1) open string field theory is equivalent to U(1) gauge theory, 
it is natural to consider string field theory analog of the 
field strength. The success of the analysis of tachyon condensation \cite{kmm}
in boundary string field theory (BSFT)\cite{w}
also suggests that we can define such an analog, because gauge field in
BSFT transforms in the same way as ordinary gauge field, and the BSFT action
is expected to be related to cubic string field theory and its superstring 
extension by some field redefinition.

As applications of our gauge invariant quantities,
we make two attempts to extract gauge invariant field strength.
Firstly, we consider string field theory counterpart 
of the coupling of Ramond-Ramond (p-1)-form and one field strength 
of Dp-brane in the effective action.
From it we extract quantities analogous to gauge invariant 
field strength and gauge field, and extend it to off-shell 
in the most naive and straightforward manner.
It is not clear from the general expression that the analog of gauge 
field transforms in the same way as ordinary one.
To make it clear and to investigate how far this quantity can be extended
to ``off-shell'', we give the component expression of this quantity up to 
level 2. We find that the field strength is not gauge invariant at level 2,
even on-shell, and the gauge field does not transform in the same way 
as ordinary gauge field.

Secondly, we extract a ``gauge invariant component'' of the string field 
equation of motion. The string field equation of motion contains 
a gauge invariant extension of that of 
ordinary gauge field i.e. $\p_\nu F^{\mu\nu}=0$. This string field theory 
counterpart has correction terms from massive fields. We compute those
terms up to level 1. From this ``gauge invariant component'' we can 
extract an analog of field strength, and at the linearized level this
component is gauge invariant even off-shell.

For definiteness, in this paper we consider one single D9-brane 
in type IIB theory. Extension of our discussion to lower dimensional 
D-branes is straightforward, but is restricted to one single D-brane.
In section 2, we introduce couplings of one closed string mode and 
one open string field which are gauge invariant 
in Berkovits' open string field theory.
Our discussion is entirely in terms of conformal field theory.
In section 3, we consider coupling of RR 8-form and one open string field,
and give explicit expression up to level 2. We show that field strength 
defined from it is not gauge invariant even on-shell.
In section 4, we define another type of ``on-shell'' gauge invariant 
which reduces to the previous ones, and
define gauge invariant components of the equation of
motion. From one of them we extract field strength and gauge field, 
and show that the field strength is gauge invariant under linearized
gauge transformation even off-shell. Section 5 contains discussions.
In the Appendix bases for expanding string fields are tabulated.

\section{``On-shell'' gauge invariants}

In this section we introduce gauge invariant quantities linear in string 
field $\Phi$, extending the argument in \cite{z,grsz,hi} for 
bosonic open string field theory. 
Our argument is based on the conformal field theory viewpoint. 
If necessary, we can rewrite the following argument in terms of oscillator
expression as in \cite{hi}, in the case of a flat background.

The action of the NS part of the open string field theory proposed 
by Berkovits \cite{b1} is
\beq
S= \frac{1}{2g^2}\left\langle\left\langle 
(e^{-\Phi}Q_B e^\Phi)(e^{-\Phi}\eta_0 e^\Phi)
 -\int_0^1 dt (e^{-t\Phi}\p_t e^{t\Phi})
 \{(e^{-t\Phi}Q_B e^{t\Phi}),(e^{-t\Phi}\eta_0 e^{t\Phi})\}
\right\rangle\right\rangle,
\eeq
where the string field $\Phi$ is Grassmann even, GSO(+), 
ghost number 0 and picture number 0 operator.
CFT correlators $\vev{\dots}$ are defined in large Hilbert space. 
For details of the definition see, for instance, \cite{bsz}.

This action is invariant under the following gauge transformation:
\beq
\delta e^\Phi=(Q_B\Omega) e^\Phi+e^\Phi(\eta_0\Omega'),
\label{gtl}
\eeq
where the gauge transformation parameters $\Omega$ and $\Omega'$
are Grassmann odd, GSO(+) operators. Their ghost number and picture
number are $(-1,0)$ and $(-1,1)$ respectively.
From this expression, we can compute the transformation of $\Phi$
order by order:
\bea
\delta \Phi & = & (Q_B\Omega)+(\eta_0\Omega') \nn
 & & +\Half[(Q_B\Omega),\Phi]+\Half[\Phi,(\eta_0\Omega')] \nn 
 & & +\frac{1}{12}(Q_B\Omega)\Phi^2-\frac{1}{6}\Phi(Q_B\Omega)\Phi
     +\frac{1}{12}\Phi^2(Q_B\Omega) \nn
 & & +\frac{1}{12}\Phi^2(\eta_0\Omega')-\frac{1}{6}\Phi(\eta_0\Omega')\Phi
     +\frac{1}{12}(\eta_0\Omega')\Phi^2 \nn
 & & +\dots.
\label{gtl2}
\eea
This transformation contains infinitely many terms with arbitrarily 
high power in $\Phi$.

The gauge invariant quantity $(V;\Phi)$ we consider in this paper is 
defined as:
\beq
(V;\Phi)\equiv \vev{V(0,0)\cdot 
f_1^{(1)}\circ\Phi(0)}_{\mbox{{\scriptsize disk}}},
\label{gi}
\eeq
where $\vev{\cdots}_{\mbox{{\scriptsize disk}}}$ is the CFT correlation 
function evaluated
on a unit disk with appropriate boundary condition on the boundary.
Conformal mappings $f^{(n)}_k(z)$, which are used to define 
star products, are
\beq
f^{(n)}_k(z)=e^{2\pi i(k-1)/n}\left(\frac{1+iz}{1-iz}\right)^{2/n}.
\eeq
$V(z,\bar{z})$ is a closed string vertex operator that satisfies
the following conditions:
\begin{itemize}
\item $[Q_B,V(z,\bar{z})]=0$, i.e. $V(z,\bar{z})$ is BRST invariant.
\item $[\eta_0,V(z,\bar{z})]=0$, i.e. $V(z,\bar{z})$ is in small Hilbert space.
\item $V(z,\bar{z})$ is a dimension $(0,0)$ primary field.
\end{itemize}
Ordinary on-shell closed string vertex operators in first quantized 
formalism satisfy all of these conditions.

There is a subtlety in the definition of $(V;\Phi)$.
The mapping $f_1^{(1)}$ is singular at the center of the unit disk 
where $V$ is inserted.
Geometrically, this mapping glues the left half and the right half of the open 
string and therefore the midpoint is singular. However, in (\ref{gi}),
$f_1^{(1)}$ acts only on $\Phi$ at the point where this mapping is not 
singular. Hence we can forget that the unit disk
is formed by the gluing procedure and can evaluate (\ref{gi}) regarding 
it as merely a CFT correlation function on the unit disk. Then (\ref{gi})
is not singular and well-defined.

We may have to take conformal transformations of $(V;\Phi)$, which is singular
at the point where $V$ is inserted. For such cases, we can define $(V;\Phi)$ 
by taking a limit:
\beq
(V;\Phi)\equiv\lim_{\epsilon\to 0}
\vev{V(\epsilon,\bar{\epsilon})\cdot f_1^{(1)}\circ
\Phi(0)}_{\mbox{{\scriptsize disk}}}.
\eeq

Since the gauge transformation of $\Phi$ is given implicitly by (\ref{gtl})
and is quite different from the bosonic string field theory counterpart 
$\delta\Phi=Q_B\Omega+\Phi*\Omega-\Omega*\Phi$, 
at first glance it is not clear that $(V;\Phi)$ is gauge invariant.
However, for proving gauge invariance it is sufficient to show the following
relation:
\beq
(V;A*B)=(V;B*A).
\label{commu}
\eeq
Since no explicit explanation for this relation in CFT viewpoint has not been
given in the literature,
we will give one shortly. In fact this is valid for any dimension (0,0) 
primary operator $V$, not necessarily BRST invariant or in small Hilbert 
space. In \cite{hi} this has been proved for the case where $V$ is a
tachyon vertex operator in terms of oscillator expression in flat background.

Now let us proceed assuming this relation.
Then the invariance of $(V;\Phi)$ can be proved as follows.
In general $\delta\Phi$ is not equal to $e^{-\Phi}\delta e^{\Phi}$, but 
thanks to (\ref{commu}), $(V;\delta\Phi)=(V;e^{-\Phi}\delta e^{\Phi})$.
Then plugging (\ref{gtl}) into this,
\bea
(V;\delta\Phi) & = & (V; Q_B\Omega+\eta_0\Omega').
\eea
So far we did not use the fact that $V$ is BRST invariant and in small 
Hilbert space, and by using it, we obtain
\beq
(V;\delta\Phi)=\vev{\left[Q_B(V(0)\cdot f_1^{(1)}\circ\Omega(0))
+\eta_0(V(0)\cdot f_1^{(1)}\circ\Omega'(0))\right]}=0.
\eeq

We now make some comments on the properties of $(V;\Phi)$.
\begin{itemize}
\item $(V;\Phi)$ is ``on-shell'' in the following sense.
 For example, in flat background momentum $q$ of $V$ satisfies the on-shell 
 condition $q^2=\mbox{const.}$, because of the BRST invariance or the 
 condition of
 (0,0) conformal dimension. Therefore, by momentum conservation $q+k=0$,
 the momentum $k$ of $\Phi$ also satisfies $k^2=\mbox{const.}$
\item $(V;\Phi)$ is linear in $\Phi$. In terms of component expression of
 the string field, it is surprising that all the nonlinear terms in the gauge
 transformation of $\Phi$ cancel.
\item Since $V$ does not have $\xi_0$, $\Phi$ has to have $\xi_0$ to obtain
 a nonzero contribution. Therefore we can concentrate on only those operators
 that survive the $\xi_0\Phi=0$ gauge condition. 
\item An obvious physical interpretation of $(V;\Phi)$ is that these 
 quantities represent couplings of one on-shell closed string mode and 
 one open string mode.\cite{z,grsz,hi}
\end{itemize}

\begin{figure}[htdp]
\begin{center}
\leavevmode
\epsfxsize=140mm
\epsfbox{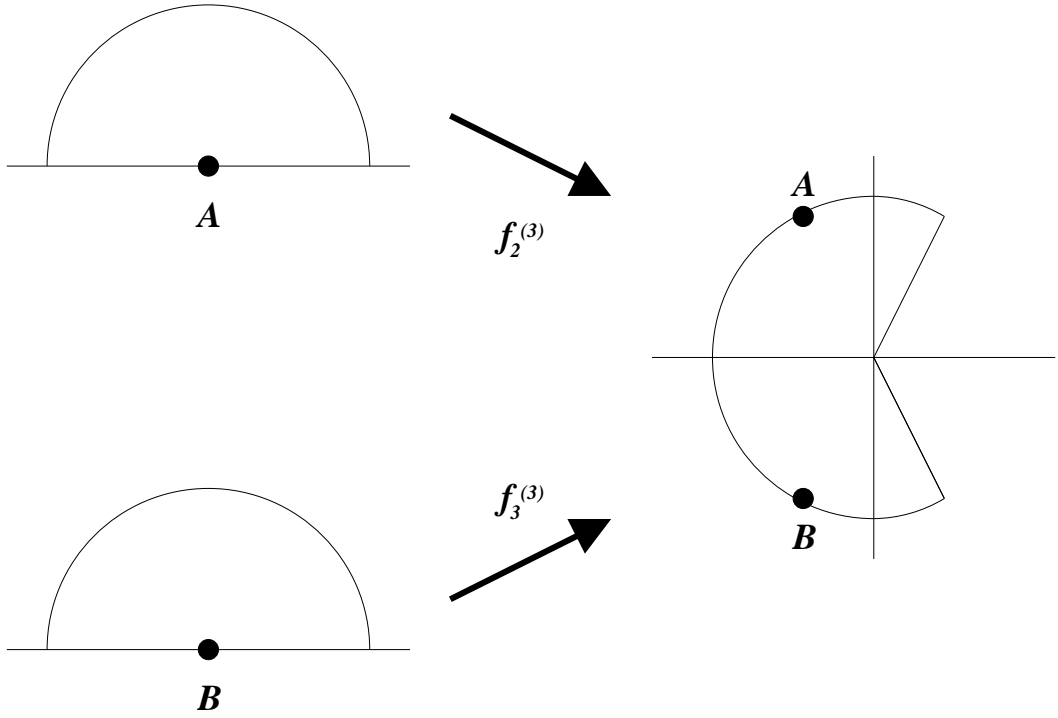}
\caption{Worldsheets of two open strings are glued together.}
\label{fig1}
\end{center}
\end{figure}

\begin{figure}[htdp]
\begin{center}
\leavevmode
\epsfxsize=140mm
\epsfbox{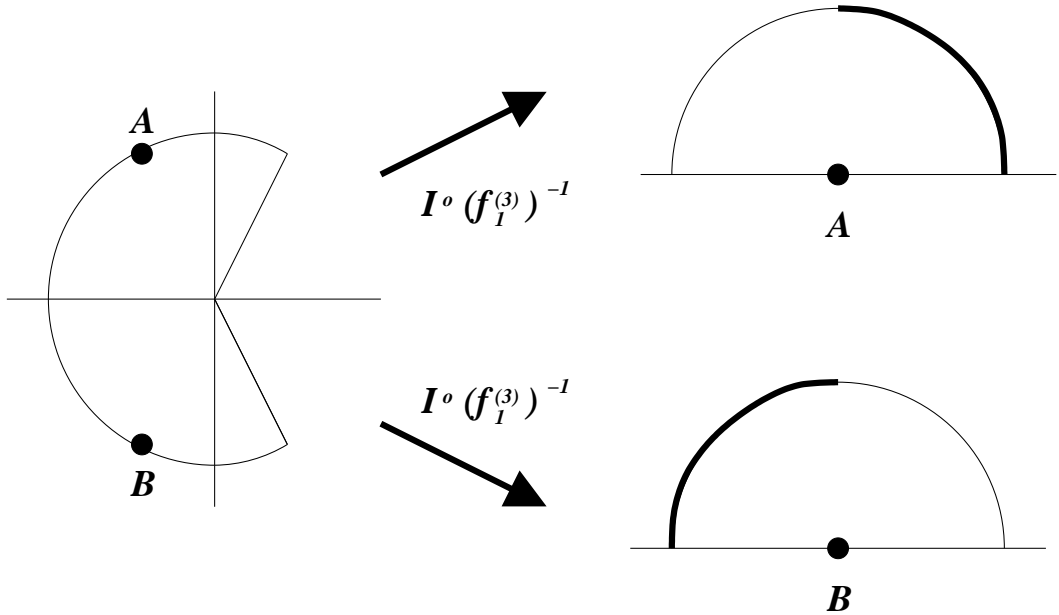}
\caption{The sector is mapped to two copies of half disk.
 Identified parts are shown by bold lines.}
\label{fig2}
\end{center}
\end{figure}

\begin{figure}[htdp]
\begin{center}
\leavevmode
\epsfxsize=140mm
\epsfbox{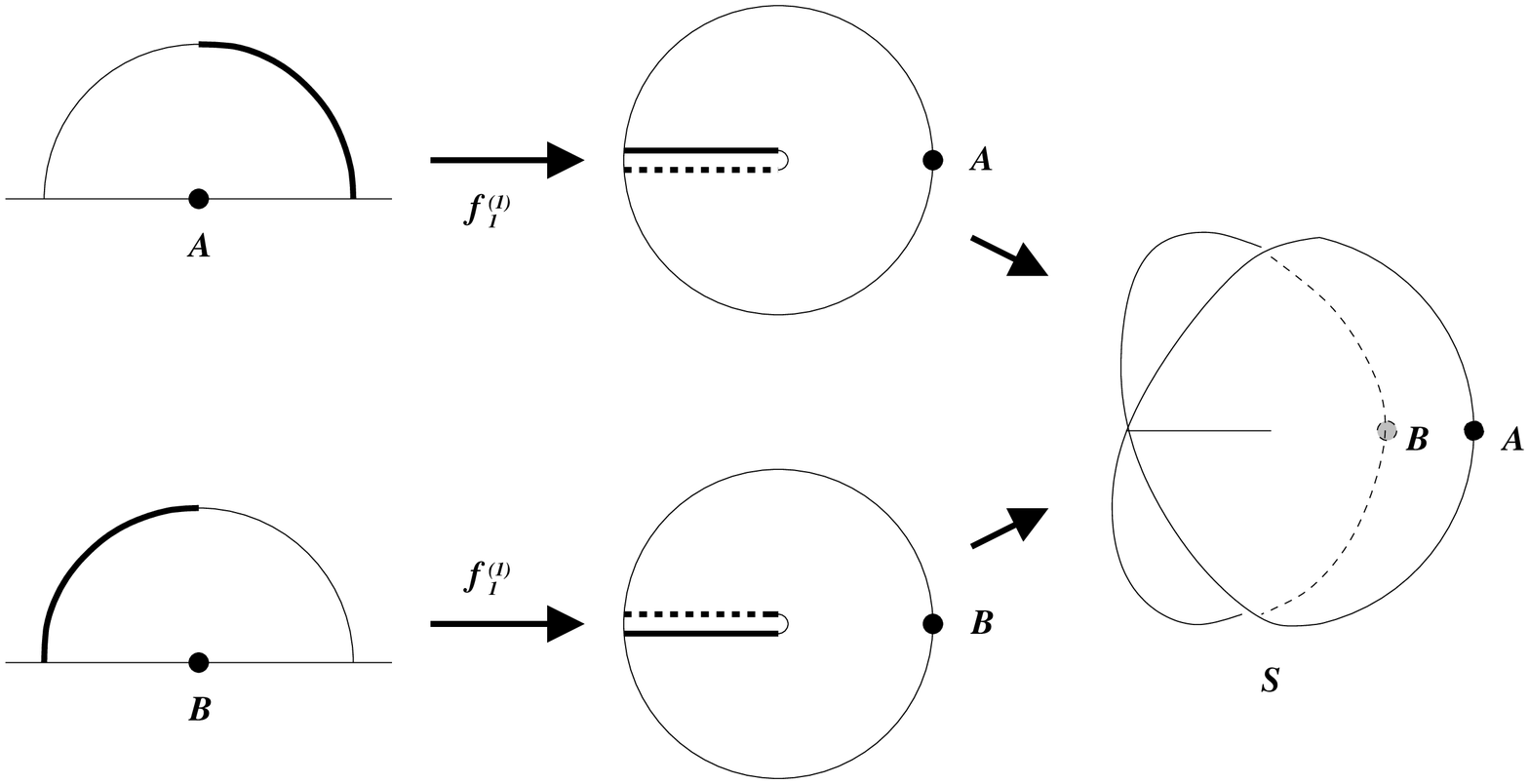}
\caption{Two half-disks are mapped to one ``double-valued'' 
unit disk $S$. Identified parts are shown by bold and dashed lines.
}
\label{fig3}
\end{center}
\end{figure}

Now we give an explanation for the relation (\ref{commu}) in terms of CFT.
We assume that $V$ is a dimension (0,0) primary operator.
Comparing the following two relations,
\bea
\langle\langle A*B*C \rangle\rangle
& = & \vev{f^{(3)}_1\circ A(0)\cdot
f^{(3)}_2\circ B(0)\cdot f^{(3)}_3\circ C(0)}_{\mbox{{\scriptsize disk}}} \nn
 & = & \vev{A(0)\cdot (f^{(3)}_1)^{-1}\circ
\left[f^{(3)}_2\circ B(0)\cdot f^{(3)}_3\circ C(0)\right]} \nn
 & = & \vev{f^{(2)}_1\circ A(0)\cdot f^{(2)}_1\circ (f^{(3)}_1)^{-1}\circ
\left[f^{(3)}_2\circ B(0)\cdot f^{(3)}_3\circ C(0)\right]} \nn
 & = & \vev{f^{(2)}_1\circ A(0)\cdot 
f^{(2)}_2\circ I\circ (f^{(3)}_1)^{-1}\circ
\left[f^{(3)}_2\circ B(0)\cdot f^{(3)}_3\circ C(0)\right]},
\\
\langle\langle A*(B*C) \rangle\rangle
& = & \vev{f^{(2)}_1\circ A(0)\cdot 
f^{(2)}_2\circ(B*C)(0)}_{\mbox{{\scriptsize disk}}},
\eea
we obtain the following CFT expression of $(V;A*B)$.
\bea
(V;A*B) & = & \lim_{\epsilon\to 0}
 \vev{V(\epsilon,\bar{\epsilon})\cdot f^{(1)}_1\circ I\circ (f^{(3)}_1)^{-1}
 \circ\left[f^{(3)}_2\circ A(0)
 \cdot f^{(3)}_3\circ B(0)\right]}_S \nn
 & = & \lim_{\epsilon\to 0}\vev{V(\epsilon,\bar{\epsilon})
 \cdot h\circ f^{(1)}_1\circ I\circ (f^{(3)}_1)^{-1}
 \circ\left[f^{(3)}_2\circ A(0)
 \cdot f^{(3)}_3\circ B(0)\right]}_{\mbox{{\scriptsize disk}}},
\label{cfte}
\eea
where $I(z)=-z^{-1}$ and $h(z)=z^{1/2}$.
Let us trace the successive actions of conformal mappings in the 
above expression.
By $f^{(3)}_2$ on $A$ and $f^{(3)}_3$ on $B$, worldsheets of 
two open strings are glued together to form a sector with the central 
angle of $4\pi/3$ (fig. \ref{fig1}). 
Then by $I\circ (f^{(3)}_1)^{-1}$ it is mapped to two copies of half-disk
with the left half arc of one copy and the right half arc of the other 
identified (fig. \ref{fig2}).
Next by $f^{(1)}_1$ they are mapped to two copies
of unit disk. These two disks have cuts along the negative part of the real
axis, and when we go across one of them, we jump to the other.
Thus we get the Riemann surface $S$ in fig. \ref{fig3}.
Finally by $h$, $(V;A*B)$ becomes a correlation
function on a unit disk. Note that $V$ is invariant under $h$. 
At this stage we can take the limit $\epsilon\to 0$ safely and 
obtain the following nonsingular result:
\bea
(V;A*B) & = &
 \vev{V(0,0)
 \cdot f^{(2)}_1\circ A(0)
 \cdot f^{(2)}_2\circ B(0)}_{\mbox{{\scriptsize disk}}}.
\eea

We can easily see that a $\pi$ rotation exchanges the positions of $A$ 
and $B$, and leaves $V(0,0)$ invariant. (If $A$ and $B$ are Grassmann odd
we have an extra sign factor.) Thus we have established relation 
(\ref{commu}). 

By extending the above argument, we obtain
\beq
(V;\Phi_1*\Phi_2*\dots*\Phi_n) =
 \vev{V(0,0)
 \cdot f^{(n)}_1\circ \Phi_1(0)
 \cdot f^{(n)}_2\circ \Phi_2(0)\dots
 f^{(n)}_n\circ \Phi_n(0)}_{\mbox{{\scriptsize disk}}}
\eeq
From this expression we see that $(V;\Phi_1*\Phi_2*\dots*\Phi_n)$ is
invariant under a cyclic exchange of $\Phi_1$, $\Phi_2,\dots$ and $\Phi_n$.

Let us give a comment on the case where $V$ is a dimension $(h_L,h_R)$
primary operator. In this case $V$ is not invariant under $h$ and
obtains the factor $(\epsilon/2)^{-h_L/2}(\bar{\epsilon}/2)^{-h_R/2}$.
In general this factor is divergent or vanishing, and makes 
the second expression of (\ref{cfte}) ill-defined.

\section{Gauge invariant RR coupling and field strength}

In this section we investigate the coupling of a string field and RR field
as an application of the gauge invariant quantities defined in the
previous section. From it we extract a string field theory analog 
of gauge invariant field strength of ordinary U(1) gauge theory
up to level 2. We will see that this field strength is not gauge 
invariant, even on-shell.

We consider one single type IIB D9-brane in a flat background, and
take RR 8-form vertex operator $V_{RR8}$ as $V$.
Then $(V_{RR8};\Phi)$ is interpreted as the coupling of RR 8-form and one
open string field. We can expect that this
corresponds to Chern--Simons term $\int C^{(8)}\wedge\wt{F}
=-\int F^{(9)}\wedge\wt{A}$ in the effective action, where $\wt{A}$ is the
ordinary gauge field on the D-brane, and $\wt{F}$ is its field strength.
Therefore we can extract a string field theory analog of $\wt{F}$
by computing $(V_{RR8};\Phi)$.

The lowest level component of $\Phi$ is equal to
$\int\frac{d^{10}k}{(2\pi)^{10}}[A_\mu(k)\xi c\psi^\mu e^{-\phi}e^{ikX}
+B(k)\xi\p\xi c\p ce^{-2\phi}e^{ik\cdot X}]$.
In lowest order calculation, $A_\mu$ corresponds to $\wt{A}_\mu$ \cite{bs},
and the lowest order term of 
$(V_{RR8};\Phi)$ should be proportional to $F_{\mu\nu}$, the field 
strength of $A_\mu$.
It is easy to see that this is indeed the case: 
$(V_{RR8};\Phi)\propto\vev{V_{RR8}(0,0) \int\frac{d^{10}k}{(2\pi)^{10}}
A_\mu(k)\xi c\psi^\mu e^{-\phi}e^{ikX}(1)}_{\mbox{{\scriptsize disk}}}
+\dots$ and noting that $c\psi^\mu e^{-\phi}e^{ikX}$ is the vertex operator 
corresponding to the gauge field in first quantized formalism,
this CFT correlator represents the coupling of RR 8-form and gauge 
field on D-brane in first quantized formalism, and gives 
$C^{(8)}\wedge F$.

Higher order terms distort the equality of $A_\mu$ and $\wt{A}_\mu$,
as can be seen from the gauge transformation law. $\wt{A}_\mu$ is expected
to transform as $\delta\wt{A}_\mu=\p_\mu\wt{\lambda}$, while the 
transformation of $A_\mu$ contains terms of arbitrarily high order in 
infinitely many modes of $\Phi$. Therefore the string theory counterpart 
of U(1) field strength has correction terms to $F$.
$(V_{RR8};\Phi)$ gives a gauge invariant 
extension of $C^{(8)}\wedge F$, and can be expected to give some 
information on the correction terms.

In the following we compute $(V_{RR8};\Phi)$ order by order, up to level 2.
The RR $p$-form vertex operator is
\bea
V_{RR p}(z,\bar{z}) & = & F^{(p+1)}_{\mu_1\dots\mu_{p+1}}(-q)
 (C\Gamma^{\mu_1\dots\mu_{p+1}})_{AB} V_{RR}(z,\bar{z})^{AB}, \nn
V_{RR}(z,\bar{z})^{AB} & = & 
 c(z)e^{-\Half\phi(z)}S^A(z)
 \wt{c}(\bar{z})e^{-\Half\wt{\phi}(\bar{z})}\wt{S}^B(\bar{z})
 e^{-iq\cdot X}(z,\bar{z}),
\eea
where $S^A(z)$ and $\wt{S}^B(\bar{z})$ are spin operators and,
as is well known, $F^{(p+1)}_{\mu_1\dots\mu_{p+1}}(q)$ corresponds 
to field strength of the RR $p$-form.
BRST invariance of $V_{RR p}$ requires 
$F^{(p+1)}(q)_{\mu_1\dots\mu_{p+1}}
= (p+1) q_{[\mu_1}C^{(p)}(q)_{\mu_2\dots\mu_{p+1}]},\quad
 q^{\mu_1}C^{(p)}(q)_{\mu_1\mu_2\dots\mu_p}=0$ and $q^2=0$.
$V_{RR p}$ also satisfies $[\eta_0, V_{RR p}]=0$, and is a dimension (0,0)
primary field.

In order to perform an order by order calculation, let us decompose $\Phi$ and
$\Omega$ into sums of contributions of each level.
The level is defined as $L_0-\ap k^2$, i.e. the eigenvalue of $L_0$ with the
contribution of $e^{ik\cdot X}$ subtracted.
\bea
\Phi & = & \Phi_0+\Phi_1+\Phi_2+\dots \nn
\Omega & = & \Omega_0+\Omega_1+\Omega_2+\dots
\eea
where $\Phi_n$ and $\Omega_n$ are level $n$ parts of $\Phi$ and $\Omega$
respectively.

By the general argument given in the previous section, $(V_{RR8};\Phi)$
is gauge invariant.
The linear part $(Q_B\Omega)+(\eta_0\Omega')$ and the higher order part
in $(V_{RR8};\delta\Phi)$ cancel separately. Henceforth we consider 
only the linear part $\delta_0$ of the gauge transformation, i.e.
$\delta_0\Phi= Q_B\Omega$. $\eta_0\Omega'$
consists of only those without $\xi_0$, and does not give nonzero 
contribution to $(V_{RR8};\Phi)$. 
Therefore we do not consider the parameter $\Omega'$.

Since $Q_B$ does not change the level, $Q_B\Omega_n$ gives the gauge 
transformation of $\Phi_n$, and we can investigate the contribution of 
each level separately.

First we consider the level 0 contribution.
As we have already seen, 
$\Phi_0=\int\frac{d^{10}k}{(2\pi)^{10}}
[A_\mu(k)\xi c\psi^\mu e^{-\phi}e^{ikX}
+B(k)\xi\p\xi c\p ce^{-2\phi}e^{ik\cdot X}]$ gives
\beq
(V_{RR}^{AB};\Phi_0) = 
 -\frac{1}{\sqrt{2}}2^{2\ap q^2}(2i)^{\ap q^2/2}
 \left[\Gamma^\mu\Half(1+\Gamma^{10})C^{-1}\right]^{AB}A_\mu(q),
\eeq
where we did not use $q^2=0$.

To extract a quantity analogous to the gauge field, we define $a_\mu(k)$ 
as follows:
\beq
a_\mu(q) = -\sqrt{2}\cdot 2^{-2\ap q^2}(2i)^{-\ap q^2/2}
\frac{1}{16}(C\Gamma_\mu)_{AB}
(V_{RR}^{AB};\Phi)_{\mbox{{\scriptsize off-shell}}},
\eeq
where $(V_{RR}^{AB};\Phi)_{\mbox{{\scriptsize off-shell}}}$ is equal to
$(V_{RR}^{AB};\Phi)$ computed without using the on-shell condition $q^2=0$.
In other words, $a_\mu(k)$ is the gauge field obtained by naive and 
straightforward off-shell
extension of $(V_{RR8};\Phi)$. At level 0, $a_\mu(k)=A_\mu(k)$. 
Then the field strength of $a_\mu(k)$ is defined by
$f_{\mu\nu}(k)=ik_\mu a_\nu(k)-ik_\nu a_\mu(k)$.

Then $(V_{RR8};\Phi)$ is given by 
\bea
(V_{RR8};\Phi) & = &  -\frac{16}{\sqrt{2}}
 2^{2\ap q^2}(2i)^{\ap q^2/2}\epsilon^{\mu_1\dots\mu_{10}}
 F^{(9)}_{\mu_1\dots\mu_9}(-q)a_{\mu_{10}}(q) \nn
 & = &  -\frac{16i}{\sqrt{2}}
 2^{2\ap q^2}(2i)^{\ap q^2/2}\epsilon^{\mu_1\dots\mu_{10}}
 C^{(8)}_{\mu_1\dots\mu_8}(-q)f_{\mu_9\mu_{10}}(q).
\eea
Let us see the linearized gauge transformation of $a_\mu(k)$.
$\Omega_0=\int\frac{d^{10}k}{(2\pi)^{10}}
\frac{i}{\sqrt{2\ap}}\lambda(k)\xi\p\xi ce^{-2\phi}e^{ikX}$ 
gives the following gauge transformation:
\beq
\delta_0 a_\mu(k)=ik_\mu\lambda(k).
\eeq
At this level, $a_\mu$ transforms in the same way as ordinary gauge field.
Note that the gauge transformation is defined off-shell.
Therefore at this level $f_{\mu\nu}(k)$ is an off-shell gauge invariant.

Next we consider the level-1 contribution.
At this level $\Phi_1$ and $\Omega_1$ are expanded by the basis given in
the tables in the Appendix. Since $V_{RR}^{AB}$ has two $c$ ghosts and $\phi$ 
charge $-1$, and does not have $\xi$ and $\eta$, only those with $bc$ ghost 
number 1, $\phi$ charge $-1$ and $\xi\eta$ ghost number 1 and with $\xi_0$ 
give nonzero contribution to 
$(V_{RR}^{AB};\Phi_1)_{\mbox{{\scriptsize off-shell}}}$ and $a_\mu$. 
Then, the relevant basis consists of
$\xi c\psi^\mu \p e^{-\phi}e^{ik\cdot X}$,
$\xi c \p\psi^\mu e^{-\phi}e^{ik\cdot X}$,
$\xi\p c \psi^\mu e^{-\phi}e^{ik\cdot X}$,
$\xi c \psi^\mu\psi^\nu\psi^\lambda e^{-\phi}e^{ik\cdot X}$ and
$\xi c \psi^\mu e^{-\phi}:\p X^\nu e^{ik\cdot X}:$.
By computing the CFT correlators of these operators, we find that
none of them contribute to $(V_{RR8};\Phi_1)_{\mbox{{\scriptsize off-shell}}}$
and $a_\mu$.

Finally we consider level-2 contribution, and see whether $a_\mu(k)$
transforms in the same way as an ordinary gauge field.
$\Phi_2$ and $\Omega_2$ are expanded by the basis given in
the tables in the Appendix. Among them only the following components give
nonzero contributions to $(V_{RR0};\Phi_2)_{\mbox{{\scriptsize off-shell}}}$
and $a_\mu(k)$:
\bea
\Phi_2 & = & \int\frac{d^{10}k}{(2\pi)^{10}}\Bigl[
 B^{(1)}_\mu(k)
 (\xi c      \psi^\mu \p^2 e^{-\phi}       e^{ik\cdot X}) \nn
 & & +
 B^{(2)}_\mu(k)
 (\xi c      \psi^\mu :\p^2\phi e^{-\phi}: e^{ik\cdot X}) \nn
 & & +
 B^{(3)}_\mu(k)
 (\xi \p^2 c \psi^\mu e^{-\phi}            e^{ik\cdot X}) \nn
 & & +
 B^{(4)}_{\mu\nu}(k)
 (\xi c      \psi^\mu e^{-\phi}         :\p^2 X^\nu e^{ik\cdot X}:) \nn
 & & +
 B^{(5)}_\mu(k)
 (\xi c      \p^2\psi^\mu   e^{-\phi}      e^{ik\cdot X}) \nn
 & & +
 B^{(6)}_{\mu\nu\lambda}(k)
 (\xi c  :\p\psi^\mu\psi^\nu\psi^\lambda:e^{-\phi}e^{ik\cdot X}) \nn
 & & +
 B^{(7)}_\mu(k)
 (:\xi\p\xi\eta: c \psi^\mu   e^{-\phi}    e^{ik\cdot X}) \nn
 & & +
 B^{(8)}_{\mu\nu\lambda}(k)
 (\xi c    \psi^\mu  e^{-\phi} :\p X^\nu \p X^\lambda e^{ik\cdot X}:) \nn
 & & +
 B^{(9)}_\mu(k)
 (\xi:bc\p c:  \psi^\mu   e^{-\phi}       e^{ik\cdot X})
\Bigr],
\eea
where the coefficients $B^{(n)}$ have appropriate symmetry, i.e.
$B^{(6)}_{\mu\nu\lambda}$ is antisymmetric under the exchange of $\nu$ and
$\lambda$, and $B^{(8)}_{\mu\nu\lambda}$ is symmetric under the exchange of 
$\nu$ and $\lambda$.

$\Omega_2$ is expanded, and coefficients of components are defined
as follows:
\bea
\Omega_2 & = & \int\frac{d^{10}k}{(2\pi)^{10}}\Bigl[
      \epsilon^{(1)}(k)
 (\xi\p^2\xi c \p e^{-2\phi}   e^{ik\cdot X})
     +\epsilon^{(2)}(k)
 (\xi\p\xi   c \p^2 e^{-2\phi} e^{ik\cdot X}) \nn
 & & +\epsilon^{(3)}(k)
 (\xi\p\xi   c :\p^2\phi e^{-2\phi}: e^{ik\cdot X})
     +\epsilon^{(4)}(k)
 (\xi\p^3\xi c e^{-2\phi}      e^{ik\cdot X}) \nn
 & & +\epsilon^{(5)}_\mu(k)
 (\xi\p^2\xi c e^{-2\phi}    :\p X^\mu e^{ik\cdot X}:)
     +\epsilon^{(6)}_\mu(k)
 (\xi\p\xi   c \p e^{-2\phi} :\p X^\mu e^{ik\cdot X}:) \nn
 & & +\epsilon^{(7)}_{\mu\nu}(k)
 (\xi\p^2\xi c :\psi^\mu\psi^\nu: e^{-2\phi}    e^{ik\cdot X})
     +\epsilon^{(8)}_{\mu\nu}(k)
 (\xi\p\xi   c :\psi^\mu\psi^\nu: \p e^{-2\phi} e^{ik\cdot X}) \nn
 & & +\epsilon^{(9)}_\mu(k)
 (\xi\p\xi\p^2\xi c\p^2 c \psi^\mu e^{-3\phi} e^{ik\cdot X})
     +\epsilon^{(10)}_\mu(k)
 (\xi\p\xi   c e^{-2\phi}    :\p^2 X^\mu e^{ik\cdot X}:) \nn
 & & +\epsilon^{(11)}_{\mu\nu}(k)
 (\xi\p\xi   c e^{-2\phi}    :\p X^\mu\p X^\nu e^{ik\cdot X}:)
     +\epsilon^{(12)}_{\mu\nu}(k)
 (\xi\p\xi   c :\psi^\mu\psi^\nu: e^{-2\phi} :\p X^\lambda 
     e^{ik\cdot X}:) \nn
 & & +\epsilon^{(13)}_{\mu\nu}(k)
 (\xi\p\xi   c :\p\psi^\mu \psi^\nu: e^{-2\phi} e^{ik\cdot X})
     +\epsilon^{(14)}_{\mu\nu\lambda\rho}(k)
 (\xi\p\xi   c :\psi^\mu\psi^\nu\psi^\lambda\psi^\rho: e^{-2\phi} 
     e^{ik\cdot X}) \nn
 & & +\epsilon^{(15)}_\mu(k)
 (\xi \psi^\mu \p e^{-\phi} e^{ik\cdot X})
     +\epsilon^{(16)}(k)
 (\xi\p\xi \p^2 c e^{-2\phi} e^{ik\cdot X}) \nn
 & & +\epsilon^{(17)}_\mu(k)
 (\xi       :bc: \psi^\mu e^{-\phi} e^{ik\cdot X})
     +\epsilon^{(18)}_\mu(k)
 (\xi      \p\psi^\mu e^{-\phi} e^{ik\cdot X}) \nn
 & & +\epsilon^{(19)}_{\mu\nu}(k)
 (\xi \psi^\mu              e^{-\phi}:\p X^\nu e^{ik\cdot X}:)
     +\epsilon^{(20)}_{\mu\nu\lambda}(k)
 (\xi :\psi^\mu\psi^\nu\psi^\lambda: e^{-\phi} e^{ik\cdot X}) \nn
 & & +\epsilon^{(21)}_\mu(k)
 (\xi\p\xi\p^2\xi c\p c \psi^\mu \p e^{-3\phi} e^{ik\cdot X})
     +\epsilon^{(22)}_\mu(k)
 (\xi\p\xi\p^3\xi c\p c \psi^\mu    e^{-3\phi} e^{ik\cdot X}) \nn
 & & +\epsilon^{(23)}(k)
 (\xi\p\xi\p^2\xi\p^3\xi c \p c \p^2 c e^{-4\phi} e^{ik\cdot X})
     +\epsilon^{(24)}_\mu
 (\xi\p\xi\p^2\xi c \p c \p\psi^\mu e^{-3\phi} e^{ik\cdot X}) \nn
 & & +\epsilon^{(25)}_{\mu\nu\lambda}(k)
 (\xi\p\xi\p^2\xi c \p c :\psi^\mu\psi^\nu\psi^\lambda: e^{-3\phi}
     e^{ik\cdot X})
     +\epsilon^{(26)}_{\mu\nu}(k)
 (\xi\p\xi\p^2\xi c \p c \psi^\mu e^{-3\phi} :\p X^\nu e^{ik\cdot X}:) \nn
 & & +\epsilon^{(27)}(k)
 (\xi\p^2\xi \p c e^{-2\phi} e^{ik\cdot X})
     +\epsilon^{(28)}(k)
 (\xi\p\xi \p c \p e^{-2\phi} e^{ik\cdot X}) \nn
 & & +\epsilon^{(29)}(k)
 (\xi\p\xi :bc \p c: e^{-2\phi} e^{ik\cdot X})
     +\epsilon^{(30)}_\mu(k)
 (\xi\p\xi \p c e^{-2\phi} :\p X^\mu e^{ik\cdot X}:) \nn
 & & +\epsilon^{(31)}_{\mu\nu}(k)
 (\xi\p\xi \p c:\psi^\mu\psi^\nu: e^{-2\phi} e^{ik\cdot X})
     +\epsilon^{(32)}(k)
 (be^{ik\cdot X})
\Bigr].
\eea
Again $\epsilon^{(n)}$ have the appropriate symmetry.
Then we can compute $a_\mu$ and the gauge transformation of $B^{(n)}$:
\bea
a_\mu(k) & = & A_\mu(k) -B^{(1)}_\mu(k) +2B^{(3)}_\mu(k)
  +2i\ap k^\nu B^{(4)}_{\mu\nu}(k)-B^{(5)}_\mu(k) \nn
 & & 
  +3B^{(6)\;\nu}_{\nu\mu}(k)
  -B^{(7)}_\mu(k)-\Half\ap B^{(8)\;\nu}_{\mu\nu}(k),
\eea

\bea
\delta_0 B^{(1)}_\mu(k) & = &
  -6\sqrt{2\ap}k_\mu\epsilon^{(1)}(k)
  +13\sqrt{2\ap}k_\mu\epsilon^{(2)}(k)
  +\sqrt{\frac{\ap}{2}}k_\mu\epsilon^{(3)}(k)
  +3\sqrt{2\ap}k_\mu\epsilon^{(4)}(k) \nn
 & & 
  -i\sqrt{\frac{\ap}{2}}\epsilon^{(5)}_\mu(k)
  +3i\sqrt{\frac{\ap}{2}}\epsilon^{(6)}_\mu(k)
  -2\sqrt{2\ap}k^\nu\epsilon^{(7)}_{\mu\nu}(k)
  +6\sqrt{2\ap}k^\nu\epsilon^{(8)}_{\mu\nu}(k)
  +8\epsilon^{(9)}_\mu(k) \nn
 & &
  -i\sqrt{\frac{\ap}{2}}\epsilon^{(10)}_\mu(k)
  +i\sqrt{\frac{\ap}{2}}\epsilon^{(12)\nu}_{\mu\nu}(k)
  +\sqrt{\frac{\ap}{2}}k^\nu\epsilon^{(13)}_{\nu\mu}(k)
  +\epsilon^{(15)}_\mu(k)
  +\Half\epsilon^{(17)}_\mu(k) \nn
 & &
  +36\epsilon^{(21)}_\mu(k)
  -24\epsilon^{(22)}_\mu(k),
 \\
\delta_0 B^{(2)}_\mu(k) & = &
  -8\sqrt{2\ap}k_\mu\epsilon^{(1)}(k)
  +12\sqrt{2\ap}k_\mu\epsilon^{(2)}(k)
  +2\sqrt{2\ap}k_\mu\epsilon^{(3)}(k)
  +6\sqrt{2\ap}k_\mu\epsilon^{(4)}(k) \nn
 & & 
  -i\sqrt{2\ap}\epsilon^{(5)}_\mu(k)
  +2i\sqrt{2\ap}\epsilon^{(6)}_\mu(k)
  -4\sqrt{2\ap}k^\nu\epsilon^{(7)}_{\mu\nu}(k)
  +8\sqrt{2\ap}k^\nu\epsilon^{(8)}_{\mu\nu}(k)
  +12\epsilon^{(9)}_\mu(k) \nn
 & & 
  -i\sqrt{2\ap}\epsilon^{(10)}_\mu(k)
  +i\sqrt{2\ap}\epsilon^{(12)\nu}_{\mu\nu}(k)
  +\sqrt{2\ap}k^\nu\epsilon^{(13)}_{\nu\mu}(k)
  +\frac{3}{2}\epsilon^{(17)}_\mu(k) \nn
 & & 
  +48\epsilon^{(21)}_\mu(k)
  -36\epsilon^{(22)}_\mu(k),
 \\
\delta_0 B^{(3)}_\mu(k) & = &
  -2\epsilon^{(9)}_\mu(k)
  +\Half\epsilon^{(15)}_\mu(k)
  -\sqrt{2\ap}k_\mu\epsilon^{(16)} \nn
 & & 
  -\frac{3}{2}\epsilon^{(17)}_\mu(k)
  -\Half\epsilon^{(18)}_\mu(k)
  -i\frac{\ap}{2}k^\nu\epsilon^{(19)}_{\mu\nu}(k),
 \\
\delta_0 B^{(4)}_{\mu\nu}(k) & = & 
  -4i\sqrt{\frac{2}{\ap}}\eta_{\mu\nu}\epsilon^{(1)}(k)
  +2i\sqrt{\frac{2}{\ap}}\eta_{\mu\nu}\epsilon^{(2)}(k)
  +i\sqrt{\frac{2}{\ap}}\eta_{\mu\nu}\epsilon^{(3)}(k)
  +6i\sqrt{\frac{2}{\ap}}\eta_{\mu\nu}\epsilon^{(4)}(k) \nn
 & & 
  -4i\sqrt{\frac{2}{\ap}}\epsilon^{(7)}_{\mu\nu}(k)
  +4i\sqrt{\frac{2}{\ap}}\epsilon^{(8)}_{\mu\nu}(k)
  +\sqrt{2\ap}k_\mu\epsilon^{(10)}_\nu(k)
  +i\sqrt{\frac{2}{\ap}}\epsilon^{(13)}_{\nu\mu}(k) \nn
 & &
  -2ik_\nu\epsilon^{(17)}_\mu(k)
  +\epsilon^{(19)}_{\mu\nu}(k),
 \\
\delta_0 B^{(5)}_{\mu\nu}(k) & = &
   -2\sqrt{2\ap}k_\mu\epsilon^{(1)}(k)
   +\sqrt{2\ap}k_\mu\epsilon^{(2)}(k)
   +\sqrt{\frac{\ap}{2}}k_\mu\epsilon^{(3)}(k)
   +3\sqrt{2\ap}k_\mu\epsilon^{(4)}(k) \nn
 & &
   -i\sqrt{\frac{\ap}{2}}\epsilon^{(5)}_\mu(k)
   +i\sqrt{\frac{\ap}{2}}\epsilon^{(6)}_\mu(k)
   -i\sqrt{\frac{\ap}{2}}\epsilon^{(10)}_\mu(k)
   -\frac{3}{4}\epsilon^{(17)}_\mu(k)
   -\epsilon^{(18)}_\mu(k),
 \\
\delta_0 B^{(6)}_{\mu\nu\lambda}(k) & = &
    2\sqrt{2\ap}k_\mu\epsilon^{(7)}_{\nu\lambda}(k)
   -2\sqrt{2\ap}k_\mu\epsilon^{(8)}_{\nu\lambda}(k)
   -i\sqrt{\frac{\ap}{2}}\epsilon^{(12)}_{\nu\lambda\mu}(k) \nn
 & &
   +\sqrt{2\ap}\epsilon^{(13)}_{\mu[\nu}(k)k_{\lambda]}
   -\Half\eta_{\mu[\nu}\epsilon^{(17)}_{\lambda]}(k)
   +3\epsilon^{(20)}_{\mu\nu\lambda}(k),
 \\
\delta_0 B^{(7)}_\mu(k) & = &
    -2\sqrt{2\ap}k_\mu\epsilon^{(2)}(k)
    -\sqrt{2\ap}k_\mu\epsilon^{(3)}(k)
    -i\sqrt{2\ap}\epsilon^{(6)}_\mu(k)
    -4\sqrt{2\ap}k^\nu\epsilon^{(8)}_{\mu\nu}(k) \nn
 & &
    -12\epsilon^{(9)}_\mu(k)
    +i\sqrt{2\ap}\epsilon^{(10)}_\mu(k)
    -i\sqrt{2\ap}\epsilon^{(12)\nu}_{\mu\nu}(k)
    -\sqrt{2\ap}k^\nu\epsilon^{(13)}_{\nu\mu}(k)
    -\epsilon^{(17)}_\mu(k) \nn
 & &
    -36\epsilon^{(21)}_\mu(k)
    +24\epsilon^{(22)}_\mu(k),
 \\
\delta_0 B^{(8)}_{\mu\nu\lambda}(k) & = &
    2i\sqrt{\frac{2}{\ap}}\eta_{\mu(\nu}\epsilon^{(5)}_{\lambda)}(k)
   -2i\sqrt{\frac{2}{\ap}}\eta_{\mu(\nu}\epsilon^{(6)}_{\lambda)}(k)
   +\sqrt{2\ap}k_\mu\epsilon^{(11)}_{\nu\lambda}(k)
   -2i\sqrt{\frac{2}{\ap}}\epsilon^{(12)}_{\mu(\nu\lambda)}(k) \nn
 & &
   +\frac{1}{\ap}\eta_{\nu\lambda}\epsilon^{(17)}_\mu(k)
   +2i\epsilon^{(19)}_{\mu(\nu}(k)k_{\lambda)},
 \\
\delta_0 B^{(9)}_\mu(k) & = & 
   -(\ap k^2+2)\epsilon^{(17)}_\mu(k)
   -12\epsilon^{(21)}_\mu(k)
   +12\epsilon^{(22)}_\mu(k)
   +\sqrt{2\ap}k_\mu\epsilon^{(29)}(k).
\eea

By using the above expression of $a_\mu$ and $\delta_0 B^{(n)}$, we can 
calculate $\delta_0 a_\mu$.
\bea
\delta_0 a_\mu(k) & = & ik_\mu\Bigg[\lambda(k)
-16i\sqrt{2\ap}\epsilon^{(1)}(k)
+16i\sqrt{2\ap}\epsilon^{(2)}(k)
+2i\sqrt{2\ap}\epsilon^{(3)}(k)
+18i\sqrt{2\ap}\epsilon^{(4)}(k) \nn
 & & 
+(\sqrt{2\ap})^3k^\nu\epsilon^{(10)}_\nu(k)
+i\sqrt{\frac{(\ap)^3}{2}}\epsilon_\nu^{(11)\nu}(k)
+3i\sqrt{\frac{\ap}{2}}\epsilon^{(13)\;\nu}_\nu(k)
+2i\sqrt{2\ap}\epsilon^{(16)}(k) \Bigg] \nn
 & & 
-16\sqrt{2\ap}k^\nu\epsilon^{(7)}_{\nu\mu}(k)
+16\sqrt{2\ap}k^\nu\epsilon^{(8)}_{\nu\mu}(k)
+4\ap k^2\epsilon^{(17)}_\mu(k).
\eea
Unfortunately $\delta_0 a_\mu$ is not in the form of an ordinary gauge 
transformation. Terms containing $\epsilon^{(7)}_{\nu\mu}$, 
$\epsilon^{(8)}_{\nu\mu}$ and $\epsilon^{(17)}_\mu$ are not proportional
to $k_\mu$. 

The gauge transformation of $f_{\mu\nu}$ is given by
\bea
\delta_0 f_{\mu\nu}(k) & = &
-16i\sqrt{2\ap}(k_\mu k^\lambda\epsilon^{(7)}_{\lambda\nu}(k)
-k_\nu k^\lambda\epsilon^{(7)}_{\lambda\mu}(k))
+16i\sqrt{2\ap}(k_\mu k^\lambda\epsilon^{(8)}_{\lambda\nu}(k)
-k_\nu k^\lambda\epsilon^{(8)}_{\lambda\mu}(k)) \nn
& &
+4i\ap k^2(k_\mu\epsilon^{(17)}_\nu(k)
-k_\nu\epsilon^{(17)}_\mu(k)).
\eea
Even on-shell $f_{\mu\nu}$ is not invariant. Of course $(V_{RR8};\Phi)$
is invariant, thanks to $k^{\mu_1}C^{(8)}_{\mu_1\mu_2\dots\mu_8}=0$.

Thus we have seen that the field strength extracted from 
the RR coupling is not gauge invariant. Therefore it has some physical 
meaning only when it is coupled with the on-shell RR-field.
It is not immediately clear whether we can modify the definition of 
$a_\mu$ so that it transforms in the same way as ordinary gauge
field.

\section{Gauge invariant component of the equation of motion 
and field strength}

As a second application of our gauge invariant quantities,
we consider in this section we ``gauge invariant components'' of the equation 
of motion, and extract a string field theory analog of field strength
from it, up to level 1. We will find that this quantity is invariant 
under the linearized gauge transformation, even off-shell.

Let us begin with a more general consideration.
If a function $f(\Phi)$ of $\Phi$ transforms under the gauge transformation 
covariantly, i.e. if $\delta f(\Phi)=[Q_B\Omega,f(\Phi)]$ or 
$[f(\Phi),\eta_0\Omega']$, $V$ does not have to be BRST invariant or in 
small Hilbert space for gauge invariance of $(V;f(\Phi))$.
What we need is only (\ref{commu}) and therefore the only necessary condition
on $V$ is that $V$ is a dimension (0,0) primary operator.

As an example of $f(\Phi)$, we take $Q_B(e^\Phi(\eta_0 e^{-\Phi}))$.
Since $Q_B(e^\Phi(\eta_0 e^{-\Phi}))=0$ is the equation of motion derived
from the gauge invariant action, it is obvious that this transforms
covariantly, and indeed $\delta[Q_B(e^\Phi(\eta_0 e^{-\Phi}))]=
[Q_B\Omega,Q_B(e^\Phi(\eta_0 e^{-\Phi}))]$. We may also take 
$\eta_0(e^{-\Phi}(Q_B e^\Phi))$, but $(V;\eta_0(e^{-\Phi}(Q_B e^\Phi)))$
is equal to $(V;Q_B(e^\Phi(\eta_0 e^{-\Phi})))$ because of 
$\eta_0(e^{-\Phi}(Q_B e^\Phi)) =
e^{-\Phi}Q_B(e^\Phi(\eta_0 e^{-\Phi}))e^\Phi$.

In fact $(V;Q_B(e^\Phi(\eta_0 e^{-\Phi})))$ is a kind of gauge invariant
we have considered in the previous sections. This can be shown as follows:
\beq
(V;Q_B(e^\Phi(\eta_0 e^{-\Phi})))=-(Q_BV;e^\Phi(\eta_0 e^{-\Phi})).
\eeq
$Q_BV$ is a dimension (0,0) primary operator. Therefore we can apply
(\ref{commu}) to the right hand side, and we obtain
\beq
(V;Q_B(e^\Phi(\eta_0 e^{-\Phi})))=-(Q_BV;-\eta_0 \Phi)=(\eta_0 Q_B V;\Phi).
\eeq
$\eta_0 Q_B V$ is BRST invariant, in small Hilbert space,
and is a dimension (0,0) primary operator.

From this calculation we see that only the linear part of the equation of 
motion contributes to these gauge invariants. In general each coefficient 
of component operators of the basis in the linear part of the equation of
motion is not gauge invariant. Therefore our gauge invariants give 
gauge invariant linear combinations of components of the linearized 
equation of motion.
Since the linearized equation of motion is $Q_B\eta_0\Phi=0$,
at the linearized level these quantities are gauge invariant even off-shell,
where off-shell means evaluating this quantity without imposing $q^2=0$.

In free U(1) gauge theory equation of motion is $\p_\nu \wt{F}^{\mu\nu}=0$,
and the left hand side is gauge invariant. Let us consider a string
field theory counterpart that reduces to $\p_\nu \wt{F}^{\mu\nu}$
in lowest order, and extract a string field theory counterpart of 
$\wt{F}^{\mu\nu}$ from it. We choose $V^\mu(z,\bar{z})=\xi(z)c(z)\psi^\mu(z)
e^{-\phi}(z)e^{-2iq\cdot X}(z)$
as $V$, where the factor $e^{-2iq\cdot X}(z)$ has only left moving part. 
This is not an ordinary closed string vertex operator, but is 
a dimension (0,0) primary operator if $q^2=0$.
We can see that this choice gives $\p_\nu \wt{F}^{\mu\nu}$ for 
the level zero component $\Phi_0$. The linear part of the equation of 
motion for $\Phi_0$ is
\beq
Q_B\eta_0\Phi_0=\int\frac{d^{10}k}{(2\pi)^{10}}\left[
c\p c \psi^\mu e^{-\phi} e^{ik\cdot X}
 [-\ap k^2 A_\mu(k)-\sqrt{2\ap}k_\mu B(k)]
+\eta c e^{ik\cdot X} 
 [\sqrt{2\ap}k_\mu A_\mu(k)+2B(k)]
\right].
\eeq
The coefficient of the second component in the right hand side
gives an algebraic equation of motion for the auxiliary field $B(k)$.
By using it, we obtain $k_\nu(ik^\mu A^\nu(k)-ik^\nu A^\mu(k))=
k_\nu F^{\mu\nu}(k)=0$ from the first component. 
This is the ordinary equation of motion of free U(1) gauge theory.
Then
\bea
(V^\mu;Q_B\eta_0\Phi_0)_{\mbox{{\scriptsize off-shell}}}
 & = & -(i)^{2\ap q^2}2^{\ap q^2}(-2i)^{\ap q^2}
 \ap\left[q^2A^\mu(q)+\sqrt{\frac{2}{\ap}}q^\mu B(q)\right] \nn
 & = & -(i)^{2\ap q^2}2^{\ap q^2}(-2i)^{\ap q^2}
 \ap iq_\nu[iq^\mu A^\nu(q)-iq^\nu A^\mu(q)],
\eea
where we keep $V^\mu$ off-shell. 

Next we calculate the contribution of level-1 part $\Phi_1$.
Only operators with $\xi_0$ contribute to the linearized equation of motion,
and their coefficients are defined as follows:
\bea
\Phi_1
 & = & \int\frac{d^{10}k}{(2\pi)^{10}}\Bigl[
 D^{(1)}_\mu(k)
 (\xi      c       \psi^\mu   \p e^{-\phi} e^{ik\cdot X}) \nn
 & & +D^{(2)}(k)
 (\xi\p\xi c\p^2 c            e^{-2\phi}    e^{ik\cdot X}) \nn
 & & +D^{(3)}(k)
 (\xi\eta                                   e^{ik\cdot X}) \nn
 & & +D^{(4)}_\mu(k)
 (\xi      c       \p\psi^\mu e^{-\phi}     e^{ik\cdot X}) \nn
 & & +D^{(5)}_{\mu\nu\lambda}(k)
 (\xi      c  :\psi^\mu\psi^\nu\psi^\lambda: e^{-\phi} e^{ik\cdot X}) \nn
 & & +D^{(6)}_{\mu\nu}(k)
 (\xi      c       \psi^\mu   e^{-\phi}    :\p X^\nu e^{ik\cdot X}:) \nn
 & & +D^{(7)}(k)
 (\xi\p^2\xi c\p c            e^{-2\phi}    e^{ik\cdot X}) \nn
 & & +D^{(8)}(k)
 (\xi\p\xi c\p c              \p e^{-2\phi} e^{ik\cdot X}) \nn
 & & +D^{(9)}_{\mu\nu}(k)
 (\xi\p\xi c\p c  :\psi^\mu\psi^\nu: e^{-2\phi}   e^{ik\cdot X}) \nn
 & & +D^{(10)}_\mu(k)
 (\xi\p\xi c\p c              e^{-2\phi}   :\p X^\mu e^{ik\cdot X}:) \nn
 & & +D^{(11)}_\mu(k)
 (\xi     \p c     \psi^\mu   e^{-\phi}     e^{ik\cdot X}) \nn
 & & +(\mbox{no $\xi_0$ part}) \Bigr].
\eea
The linearized equation of motion is given as follows:
\bea
Q_B\eta_0\Phi_1 & = & \int\frac{d^{10}k}{(2\pi)^{10}}\Biggl[
 (c\p c \psi^\mu \p e^{-\phi} e^{ik\cdot X}) \nn
 & & \times\Biggl[
 -(\ap k^2+1)D^{(1)}_\mu(k)
 +2\sqrt{2\ap}k_\mu D^{(7)}(k)
 -4\sqrt{2\ap}k_\mu D^{(8)}(k) \nn
 & &
 -2\sqrt{2\ap}k^\nu D^{(9)}_{\mu\nu}(k)
 -i\sqrt{\frac{\ap}{2}}D^{(10)}_\mu(k)
 +D^{(11)}_\mu(k)
 \Biggr] \nn
 & & +(\p\xi c\p c\p^2 c e^{-2\phi} e^{ik\cdot X}) \nn
 & & \times\Biggl[
 (\ap k^2+1)D^{(2)}(k)
 -D^{(7)}(k)
 +i\frac{\ap}{2}k^\mu D^{(10)}_\mu(k)
 \Biggr] \nn
 & & +(\eta\p c e^{ik\cdot X}) \nn
 & & \times\Biggl[-(\ap k^2+1)D^{(3)}(k)
 -6 D^{(7)}(k)
 +8 D^{(8)}(k)
 +\sqrt{2\ap}k^\mu D^{(11)}_\mu(k)
 \Biggr] \nn
 & & +(c\p c \p\psi^\mu e^{-\phi} e^{ik\cdot X}) \nn
 & & \times\Biggl[
 -(\ap k^2+1)D^{(4)}_\mu(k)
 -2\sqrt{2\ap}k_\mu D^{(7)}(k)
 +2\sqrt{2\ap}k_\mu D^{(8)}(k)
 +i\sqrt{\frac{\ap}{2}}D^{(10)}_\mu(k)
 +D^{(11)}_\mu(k)
 \Biggr] \nn
 & & +(c\p c:\psi^\mu\psi^\nu\psi^\lambda: e^{-\phi} e^{ik\cdot X}) \nn
 & & \times\Biggl[
 -(\ap k^2+1)D^{(5)}_{\mu\nu\lambda}(k)
 -\sqrt{2\ap}k_{[\mu} D^{(9)}_{\nu\lambda]}(k)
 \Biggr] \nn
 & & +(c\p c \psi^\mu e^{-\phi} :\p X^\nu e^{ik\cdot X}:) \nn
 & & \times\Biggl[
 -(\ap k^2+1)D^{(6)}_{\mu\nu}(k)
 -2i\sqrt{\frac{2}{\ap}}\eta_{\mu\nu}D^{(7)}(k)
 +2i\sqrt{\frac{2}{\ap}}\eta_{\mu\nu}D^{(8)}(k)
 +2i\sqrt{\frac{2}{\ap}}D^{(9)}_{\mu\nu}(k) \nn
 & &
 -\sqrt{2\ap}k_\mu D^{(10)}_\nu(k)
 +2ik_\nu D^{(11)}_\mu(k)
 \Biggr] \nn
 & & +(c\p^2 c \psi^\mu e^{-\phi} e^{ik\cdot X}) \nn
 & & \times\Biggl[
 -\Half D^{(1)}_\mu(k)
 -\sqrt{2\ap}k_\mu D^{(2)}(k)
 -\Half D^{(4)}_\mu(k)
 +i\frac{\ap}{2}k^\nu D^{(6)}_{\mu\nu}(k)
 +D^{(11)}_\mu(k)
 \Biggr] \nn
 & & +(\p\eta c e^{ik\cdot X}) \nn
 & & \times\Biggl[
 -\sqrt{2\ap}k^\mu D^{(1)}_\mu(k)
 +6D^{(2)}(k)
 -D^{(3)}(k)
 +\sqrt{2\ap}k^\mu D^{(4)}_\mu(k)
 -i\sqrt{\frac{\ap}{2}}D^{(6)\mu}_\mu(k) \nn
 & &
 +8D^{(7)}(k)
 -12D^{(8)}(k)
 \Biggr] \nn
 & & +(\eta c \p\phi e^{ik\cdot X}) \nn
 & & \times\Biggl[
 -2\sqrt{2\ap}k^\mu D^{(1)}_\mu(k)
 +8D^{(2)}(k)
 +\sqrt{2\ap}k^\mu D^{(4)}_\mu(k)
 -i\sqrt{\frac{\ap}{2}}D^{(6)\mu}_\mu(k) \nn
 & &
 +12D^{(7)}(k)
 -20D^{(8)}(k)
 \Biggr] \nn
 & & +(\eta c :\psi^\mu\psi^\nu: e^{ik\cdot X}) \nn
 & & \times\Biggl[
 -\sqrt{2\ap}k_{[\mu} D^{(1)}_{\nu]}(k)
 +3\sqrt{2\ap}k^\lambda D^{(5)}_{\mu\nu\lambda}(k)
 +i\sqrt{\frac{\ap}{2}}D^{(6)}_{[\mu\nu]}(k)
 +2D^{(9)}_{\mu\nu}(k)
 \Biggr] \nn
 & & +(\eta c :\p X^\mu e^{ik\cdot X}:) \nn
 & & \times\Biggl[
 -i\sqrt{\frac{2}{\ap}}D^{(1)}_\mu(k)
 -2ik_\mu D^{(3)}(k)
 +i\sqrt{\frac{2}{\ap}}D^{(4)}_\mu(k)
 +\sqrt{2\ap}k^\nu D^{(6)}_{\nu\mu}(k)
 +2D^{(10)}_\mu(k)
 \Biggr]
 \Biggr].
\eea
The first six components of the above equation have Klein--Gordon operator
$(\ap k^2+1)$, and the last five components do not. Therefore the first
six are dynamical equations of motion and the rest is algebraic equations
for auxiliary fields $D^{(7)}$, $D^{(8)}$, $D^{(9)}_{\mu\nu}$, 
$D^{(10)}_\mu$, and $D^{(11)}_\mu$. This can also be seen from the fact 
that operators corresponding to these auxiliary fields contain $c_0$, and
can be gauged away by taking the gauge $b_0\Phi=0$.

Computation of $(V^\mu;Q_B\eta_0\Phi_1)$ is straightforward.
\bea
(V_\mu;Q_B\eta_0\Phi_1) & = & -i(i)^{2\ap q^2}2^{\ap q^2}(-2i)^{\ap q^2}
 \Bigl[-2\ap q^2 D^{(1)}_\mu(q)
  +4\sqrt{2\ap}q_\mu D^{(2)}(q)
  -2\ap q^2 D^{(4)}_\mu(q) \nn
 & &
  +2i(\ap)^2 q^2 q^\nu D^{(6)}_{\mu\nu}(q)
  -4\sqrt{2\ap}q_\mu D^{(7)}(q)
  +2i\sqrt{2\ap}\ap q_\mu q^\nu D^{(10)}_\nu(q) \nn
 & &
  +4\ap q^2 D^{(11)}_\mu(q)
\Bigr].
\eea
Eliminating auxiliary fields $D^{(7)}$, $D^{(10)}_\mu$, and $D^{(11)}_\mu$
by algebraic equations of motion, we obtain
\bea
(V_\mu;Q_B\eta_0\Phi_1) & = & 
 -i(i)^{2\ap q^2}2^{\ap q^2}(-2i)^{\ap q^2}\ap q^2
 \Bigl[8\sqrt{2\ap}(\ap q^2+1)q_\mu D^{(2)}(q)
 -4\sqrt{2\ap}(\ap q^2+1)q_\mu D^{(3)}(q) \nn
 & &
 +4\ap (2q_\mu q^\nu D^{(4)}_\nu(q) +q^2 D^{(4)}_\mu(q)) \nn
 & & 
 +2i\ap q^\nu(D^{(6)}_{\mu\nu}(q)+D^{(6)}_{\nu\mu}(q))
 -2i\ap q_\mu D^{(6)\nu}_\nu(q)
 -4i(\ap)^2 q^\mu q^\nu q^\lambda D^{(6)}_{\nu\lambda}(q)
\Bigr].
\eea
Thus, up to this order, we get the following quantity:
\bea
 & & iq_\nu [iq_\mu A^\nu(q)-iq^\nu A_\mu(q)] \nn
 & & +q^2
 \Bigl[8\sqrt{2\ap}(\ap q^2+1)q_\mu D^{(2)}(q)
 -4\sqrt{2\ap}(\ap q^2+1)q_\mu D^{(3)}(q) \nn
 & &
 +4\ap (2q_\mu q^\nu D^{(4)}_\nu(q) +q^2 D^{(4)}_\mu(q)) \nn
 & & 
 +2i\ap q^\nu(D^{(6)}_{\mu\nu}(q)+D^{(6)}_{\nu\mu}(q))
 -2i\ap q_\mu D^{(6)\nu}_\nu(q)
 -4i(\ap)^2 q^\mu q^\nu q^\lambda D^{(6)}_{\nu\lambda}(q)
\Bigr]+\dots.
\eea
This is gauge invariant if $q^2=0$, and as we have already noticed, 
this is invariant under a linearized gauge transformation even off-shell.
This is interpreted as an extension of $q_\nu F^{\mu\nu}(q)$. However
the contribution from the level-1 part is not in the form of 
$q_\nu (\dots)^{[\mu\nu]}$. It is thus not possible to extract
the counterpart of $F^{\mu\nu}(q)$ from this quantity. 

Therefore we want to make a better choice for $V$. To give a gauge invariant 
in the form of $q_\nu(\dots)^{[\mu\nu]}$, $(V^\mu;Q_B\eta_0\Phi)$ 
has to satisfy 
$q_\mu(V^\mu;Q_B\eta_0\Phi)=0$. Note the following equation:
\bea
Q_B[\xi\p\xi ce^{-2\phi}e^{-2iq\cdot X}]
 & = & q_\mu V'^\mu-e^{-2iq\cdot X} \label{brstexact}, \\
V'^\mu & = & V^\mu
 +\sqrt{\frac{\ap}{2}}q^\mu\xi\p\xi c\p c e^{-2\phi}e^{-2iq\cdot X}.
\eea
This shows that if we add 
$\sqrt{\frac{\ap}{2}}q^\mu(\xi\p\xi c\p c e^{-2\phi}e^{-2iq\cdot X};
Q_B\eta_0\Phi)$ 
to $(V^\mu;Q_B\eta_0\Phi)$, we obtain a gauge invariant in the form of 
$q_\nu(\dots)^{[\mu\nu]}$. This is because the second term in the 
right hand side of (\ref{brstexact}) does not contribute since 
there is no zero mode of $\xi$ in the correlator, and when $V'^\mu$
is contracted with $q_\mu$ the whole thing in the correlator becomes
BRST exact.

The component expression of $(V'^\mu;Q_B\eta_0\Phi)$ up to level 1 is 
given by
\bea
(V'_\mu;Q_B\eta_0\Phi) & = & (i)^{2\ap q^2}2^{\ap q^2}(-2i)^{\ap q^2}\ap
 \Bigl[q^\nu(q_\mu A_\nu-q_\nu A_\mu)
 -2iq^\nu(q_\mu D^{(1)}_\nu-q_\nu D^{(1)}_\mu)
 \nn
 & &
 -2iq^\nu(q_\mu D^{(4)}_\nu-q_\nu D^{(4)}_\mu)
 -2\ap q^\nu q_\lambda(q_\mu D^{(6)\lambda}_\nu-q_\nu D^{(6)\lambda}_\mu)
 +4iq^\nu(q_\mu D^{(11)}_\nu-q_\nu D^{(11)}_\mu)
\Bigr] \nn
& = & (i)^{2\ap q^2}2^{\ap q^2}(-2i)^{\ap q^2}\ap
 q^\nu\Bigl[q_\mu(A_\nu-2iD^{(1)}_\nu-2iD^{(4)}_\nu
 -2\ap q_\lambda D^{(6)\lambda}_\nu+4iD^{(11)}_\nu) \nn
& & 
 -q_\nu (A_\mu-2iD^{(1)}_\mu-2iD^{(4)}_\mu-2\ap q_\lambda D^{(6)\lambda}_\mu
 +4iD^{(11)}_\mu)
\Bigr].
\eea
This is in the form of $q_\nu (\dots)^{[\mu\nu]}$. We have thus succeeded 
in extracting string field theory counterparts $f_{\mu\nu}$ and $a_\mu$
of $\wt{F}_{\mu\nu}$ and $\wt{A}_\mu$, up to this level. Note that we did
not use algebraic equations of motion for auxiliary fields to extract these
quantities.
\bea
f_{\mu\nu}(q) & = & iq_\mu a_\nu(q)-iq_\nu a_\mu(q), \nn
a_\mu(q) & = & A_\mu(q)-2iD^{(1)}_\mu(q)-2iD^{(4)}_\mu(q)
-2\ap q_\lambda D^{(6)\lambda}_\mu(q)
 +4iD^{(11)}_\mu(q).
\eea
The linearized gauge transformation of $a_\mu(q)$ is
\beq
\delta_0 a_\mu(q)= iq_\mu[\lambda(q)
-2i\sqrt{2\ap}\zeta^{(1)}(q)
-\sqrt{2\ap}\ap q^\nu\zeta^{(4)}_\nu(q)
+2i\sqrt{2\ap}\zeta^{(7)}(q)
],
\eeq
where $\zeta^{(n)}$ are defined as:
\bea
\Omega_1 & = & \int\frac{d^{10}k}{(2\pi)^{10}}[
 \zeta^{(1)}(k)
 \xi\p^2\xi c                    e^{-2\phi}    e^{ikX} \nn
 & & +\zeta^{(2)}(k)
 \xi\p\xi   c                    \p e^{-2\phi} e^{ikX} \nn
 & & +\zeta^{(3)}_{\mu\nu}(k)
 \xi\p\xi   c :\psi^\mu\psi^\nu: e^{-2\phi}    e^{ikX} \nn
 & & +\zeta^{(4)}_\mu(k)
 \xi\p\xi   c                    e^{-2\phi}   :\p X^\mu e^{ikX}: \nn
 & & +\zeta^{(5)}_\mu(k)
 \xi          \psi^\mu           e^{-\phi}     e^{ikX} \nn
 & & +\zeta^{(6)}_\mu(k)
 \xi\p\xi\p^2\xi  c\p c  \psi^\mu  e^{-3\phi}  e^{ikX} \nn
 & & +\zeta^{(7)}(k)
 \xi\p\xi   \p c                 e^{-2\phi}    e^{ikX}
],
\eea
and the linearized gauge transformation for each component is
\bea
\delta_0 D^{(1)}_\mu(k) & = &
 -2\sqrt{2\ap}k_\mu\zeta^{(1)}(k)
 +4\sqrt{2\ap}k_\mu\zeta^{(2)}(k)
 +2\sqrt{2\ap}k^\nu\zeta^{(3)}_{\mu\nu}(k)
 +i\sqrt{\frac{\ap}{2}}\zeta^{(4)}_\mu(k)
 \nn
 & & 
 -\zeta^{(5)}_\mu(k)
 +4\zeta^{(6)}_\mu(k),
 \\
\delta_0 D^{(2)}(k) & = &
 -\zeta^{(1)}(k)
 +i\frac{\ap}{2}k^\mu\zeta^{(4)}_\mu(k)
 +\zeta^{(7)}(k),
 \\
\delta_0 D^{(3)}(k) & = &
 6\zeta^{(1)}(k)
 -8\zeta^{(2)}(k)
 -\sqrt{2\ap}k^\mu\zeta^{(5)}_\mu(k) 
 +2\zeta^{(7)}(k),
 \\
\delta_0 D^{(4)}_\mu(k) & = &
 2\sqrt{2\ap}k_\mu\zeta^{(1)}(k)
 -2\sqrt{2\ap}k_\mu\zeta^{(2)}(k)
 -i\sqrt{\frac{\ap}{2}}\zeta^{(4)}_\mu(k)
 -\zeta^{(5)}_\mu(k),
 \\
\delta_0 D^{(5)}_{\mu\nu\lambda}(k) & = &
 \sqrt{2\ap}k_{[\mu}\zeta^{(3)}_{\nu\lambda]}(k),
 \\
\delta_0 D^{(6)}_{\mu\nu}(k) & = &
 2i\sqrt{\frac{2}{\ap}}\eta_{\mu\nu}\zeta^{(1)}(k)
 -2i\sqrt{\frac{2}{\ap}}\eta_{\mu\nu}\zeta^{(2)}(k)
 -2i\sqrt{\frac{2}{\ap}}\zeta^{(3)}_{\mu\nu}(k)
 \nn
 & & 
 +\sqrt{2\ap}k_\mu\zeta^{(4)}_\nu(k)
 -2ik_\nu\zeta^{(5)}_\mu(k),
 \\
\delta_0 D^{(7)}(k) & = &
 -(\ap k^2+1)\zeta^{(1)}(k)
 -\sqrt{2\ap}k^\mu\zeta^{(6)}_\mu(k)
 +\zeta^{(7)}(k),
 \\
\delta_0 D^{(8)}(k) & = &
 -(\ap k^2+1)\zeta^{(2)}(k)
 -\sqrt{2\ap}k^\mu\zeta^{(6)}_\mu(k),
 +\zeta^{(7)}(k),
 \\
\delta_0 D^{(9)}_{\mu\nu}(k) & = &
 -(\ap k^2+1)\zeta^{(3)}_{\mu\nu}(k)
 +2\sqrt{2\ap}k_{[\mu}\zeta^{(6)}_{\nu]}(k),
 \\
\delta_0 D^{(10)}_\mu(k) & = &
 -(\ap k^2+1)\zeta^{(4)}_\mu(k)
 +2i\sqrt{\frac{2}{\ap}}\zeta^{(6)}_\mu(k)
 +2ik_\mu\zeta^{(7)}(k),
 \\
\delta_0 D^{(11)}_\mu(k) & = &
 -(\ap k^2+1)\zeta^{(5)}_\mu(k)
 +2\zeta^{(6)}_\mu(k)
 +\sqrt{2\ap}k_\mu\zeta^{(7)}(k).
\eea

$V'_\mu$ is one of the simplest choices for $V$, but of course
we can take different ones. For example,
\bea
V''^\mu(z,\bar{z}) & = & \xi(z)c(z)\psi^\mu(z)e^{-\phi}(z)
 e^{-iq\cdot X}(z,\bar{z})
 \nn
 & &
 +\eta(\bar{z})\xi(z)\p\xi(z)c(z)
 e^{\phi}(\bar{z})e^{-2\phi}(z)\psi^\mu(\bar{z})e^{-iq\cdot X}(z,\bar{z})
 \nn
 & &
 +\Half\sqrt{\frac{\ap}{2}}q^\mu
 \xi(z)\p\xi(z)c(z)(\p c(z)+\p c(\bar{z}))
 e^{-2\phi}(z)e^{-iq\cdot X}(z,\bar{z})
 \nn
 & &
 -i\sqrt{\frac{2}{\ap}}
 \xi(z)\p\xi(z)c(z)c(\bar{z})e^{-2\phi}(z)
 :\p X^\mu(\bar{z})e^{-iq\cdot X}(z,\bar{z}):,
\eea
where $e^{-iq\cdot X}(z,\bar{z})$ is an ordinary momentum factor for closed 
strings, and the normal ordering of 
$:\p X^\mu(\bar{z})e^{-iq\cdot X}(z,\bar{z}):$ acts on left mover and 
right mover separately. This satisfies \\
$Q_B[\xi(z)\p\xi(z)c(z)e^{-2\phi}(z)e^{-iq\cdot X}(z,\bar{z})]=
-\sqrt{\frac{\ap}{2}}q_\mu V''^\mu(z,\bar{z})-e^{-iq\cdot X}(z,\bar{z})$.
This equation shows that $V''^\mu(z,\bar{z})$ is one of the right choices
for $V$.

We have considered only the linearized part of the gauge transformation.
Some choices for $V$ might give gauge invariant quantities under full gauge 
transformation even off-shell, and others might not.

\section{Discussion}

We have considered ``on-shell'' gauge invariants in Berkovits' open 
superstring field theory and, as an application of these
we have computed a string field theory analog of RR coupling 
$\int C^{(8)}\wedge\wt{F}$ up to level 2, and a gauge invariant 
component of the equation of motion up to level 1.
We have extracted analogs of the field strength from them.
The former is not gauge invariant even on-shell, and the corresponding
gauge field $a_\mu$ does not transform as an ordinary gauge field.
The latter is gauge invariant under full gauge transformation if on-shell,
and invariant under linearized gauge transformation even off-shell.

It seems that the latter is more promising, but obviously there are 
many open problems.
First of all, it is not clear whether it is fully invariant off-shell for 
some choices of $V$ and, if not,
whether there is a way of extending it to off-shell. 
By momentum conservation we always have some 
kind of on-shell condition if we use on-shell closed string vertex operators,
or dimension (0,0) operators.

Another unclear point is what our gauge invariant gives after integrating 
out massive modes, and how it is related to the gauge field given in 
\cite{d,cst}
and marginal deformation in \cite{sz,in}. 

The construction of field strength in the nonabelian case, 
i.e. multiple D-brane case,
is an interesting direction of extension, although our method is not 
directly applicable to this case. For RR coupling, the nonabelian
part of the field strength does not couple to the RR 8-form. More 
technically, relation (\ref{commu}) does not hold for matrix valued
string fields.

\vs{.5cm}
\noindent
{\large\bf Acknowledgments}\\[.2cm]
This work is supported by the Japan Society for the Promotion of Science 
and the Swiss National Science Foundation.

\vs{.5cm}
\noindent
{\Large\bf Appendix}\\[.2cm]
In this appendix we tabulate the bases for expanding the string field 
$\Phi$ and 
the gauge transformation parameter $\Omega$. The following two vacua are used
in the tables:
\beq
\ket{\Omega}=c_1 e^{ik\cdot X(0)}\vac,\quad
\ket{\wt{\Omega}}=e^{-\phi}(0)c_1 e^{ik\cdot X(0)}\vac,
\eeq
where $\vac$ is Fock vacuum defined by 
$\alpha^\mu_{n\geq 1}\vac=\psi^\mu_{n\geq 1/2}\vac=b_{n\geq -1}\vac
=c_{n\geq 2}\vac=\beta_{n\geq -1/2}\vac=\gamma_{n\geq 3/2}\vac=0$.
We show both oscillator expressions and operator expressions.
Oscillator expression $\ket{\mbox{osc.}}$ and the corresponding 
operator expression $v$ is related by $\ket{\mbox{osc.}}=
(\mbox{numerical factor})v(0)\vac$.

{\scriptsize

\begin{table}[htbp]
\begin{center}
\begin{tabular}{|c|r|l|} \hline
 Level & Oscillator & Operator \\
       & expression & expression \\ \hline\hline
 0 & $\xi_0\psi_{-1/2}^\mu\ket{\wt{\Omega}}$
   & $\xi c\psi^\mu e^{-\phi}e^{2ik\cdot X}$
 \\ \hline\hline
 1
   & $\xi_0\beta_{-1/2}\gamma_{-1/2}\psi_{-1/2}^\mu\ket{\wt{\Omega}}$
   & $\xi c\psi^\mu \p e^{-\phi}e^{2ik\cdot X}$ 
 \\ \cline{2-3}
   & $\xi_0\beta_{-1/2}c_{-1}\ket{\wt{\Omega}}$
   & $\xi\p\xi c\p^2 c e^{-2\phi}e^{2ik\cdot X}$ 
 \\ \cline{2-3}
   & $\xi_0\gamma_{-1/2}b_{-1}\ket{\wt{\Omega}}$
   & $:\xi\eta:e^{2ik\cdot X}$ 
 \\ \cline{2-3}
   & $\xi_0\psi_{-3/2}^\mu\ket{\wt{\Omega}}$
   & $\xi c \p\psi^\mu e^{-\phi}e^{2ik\cdot X}$ 
 \\ \cline{2-3} 
   & $\xi_0\psi_{-1/2}^\mu\psi_{-1/2}^\nu\psi_{-1/2}^\lambda\ket{\wt{\Omega}}$
   & $\xi c:\psi^\mu\psi^\nu\psi^\lambda:e^{-\phi}e^{2ik\cdot X}$ 
 \\ \cline{2-3}
   & $\xi_0\psi_{-1/2}^\mu\alpha_{-1}^\nu\ket{\wt{\Omega}}$
   & $\xi c \psi^\mu e^{-\phi}:\p X^\nu e^{2ik\cdot X}:$ 
 \\ \hline\hline
 2 & $\xi_0\psi_{-5/2}^\mu\ket{\wt{\Omega}}$
   & $\xi c \psi^\mu \p^2 e^{-\phi} e^{ikX}$
 \\ \cline{2-3}
   & $\xi_0\beta_{-3/2}\gamma_{-1/2}\psi_{-1/2}^\mu\ket{\wt{\Omega}}$
   & $(\xi c\p^2 e^{-\phi}
     -2:\eta\xi\p\xi:ce^{-\phi})\psi^\mu e^{ikX}$
 \\ \cline{2-3}
   & $\xi_0\gamma_{-3/2}\beta_{-1/2}\psi_{-1/2}^\mu\ket{\wt{\Omega}}$
   & $(\xi c(\p\phi\p\phi+\p^2\phi) e^{-\phi}
     -2:\eta\xi\p\xi:ce^{-\phi})\psi^\mu e^{ikX}$
 \\ \cline{2-3}
   & $\xi_0\psi_{-3/2}^\mu\beta_{-1/2}\gamma_{-1/2}\ket{\wt{\Omega}}$
   &
 \\ \cline{2-3}
   & $\xi_0\psi_{-3/2}^\mu\psi_{-1/2}^\nu\psi_{-1/2}^\lambda\ket{\wt{\Omega}}$ 
   & $\xi c:\p\psi^\mu\psi^\nu\psi^\lambda: e^{-\phi} e^{ikX}$
 \\ \cline{2-3}
   & $\xi_0\beta_{-3/2}c_{-1}\ket{\wt{\Omega}}$
   &
 \\ \cline{2-3}
   & $\xi_0\gamma_{-3/2}b_{-1}\ket{\wt{\Omega}}$
   &
 \\ \cline{2-3}
   & $\xi_0\psi_{-3/2}^\mu\alpha_{-1}^\nu\ket{\wt{\Omega}}$
   &
 \\ \cline{2-3}
   & $\xi_0\beta_{-1/2}\beta_{-1/2}\gamma_{-1/2}\gamma_{-1/2}
     \psi_{-1/2}^\mu\ket{\wt{\Omega}}$
   &
 \\ \cline{2-3}
   & $\xi_0\beta_{-1/2}\gamma_{-1/2}\psi_{-1/2}^\mu\psi_{-1/2}^\nu
     \psi_{-1/2}^\lambda\ket{\wt{\Omega}}$
   &
 \\ \cline{2-3}
   & $\xi_0\psi_{-1/2}^\mu\psi_{-1/2}^\nu\psi_{-1/2}^\lambda\psi_{-1/2}^\rho
     \psi_{-1/2}^\sigma\ket{\wt{\Omega}}$
   & $\xi c:\psi^\mu\psi^\nu\psi^\lambda\psi^\rho\psi^\sigma:
     e^{-\phi} e^{ikX}$
 \\ \cline{2-3}
   & $\xi_0\beta_{-1/2}\beta_{-1/2}\gamma_{-1/2}c_{-1}
     \ket{\wt{\Omega}}$
   &
 \\ \cline{2-3}
   & $\xi_0\beta_{-1/2}\gamma_{-1/2}\gamma_{-1/2}b_{-1}
     \ket{\wt{\Omega}}$
   &
 \\ \cline{2-3}
   & $\xi_0\beta_{-1/2}\gamma_{-1/2}
     \psi_{-1/2}^\mu\alpha_{-1}^\nu\ket{\wt{\Omega}}$
   &
 \\ \cline{2-3}
   & $\xi_0\beta_{-1/2}\psi_{-1/2}^\mu\psi_{-1/2}^\nu c_{-1}
     \ket{\wt{\Omega}}$
   &
 \\ \cline{2-3}
   & $\xi_0\gamma_{-1/2}\psi_{-1/2}^\mu\psi_{-1/2}^\nu b_{-1}
     \ket{\wt{\Omega}}$
   &
 \\ \cline{2-3}
   & $\xi_0\psi_{-1/2}^\mu\psi_{-1/2}^\nu\psi_{-1/2}^\lambda\alpha_{-1}^\rho
     \ket{\wt{\Omega}}$
   &
 \\ \cline{2-3}
   & $\xi_0\beta_{-1/2}c_{-2}\ket{\wt{\Omega}}$
   &
 \\ \cline{2-3}
   & $\xi_0\gamma_{-1/2}b_{-2}\ket{\wt{\Omega}}$
   &
 \\ \cline{2-3}
   & $\xi_0\psi_{-1/2}^\mu\alpha_{-2}^\nu\ket{\wt{\Omega}}$
   & $\xi c \psi^\mu e^{-\phi} :\p^2 X^\nu e^{ikX}:$
 \\ \cline{2-3}
   & $\xi_0\beta_{-1/2}c_{-1}\alpha_{-1}^\mu\ket{\wt{\Omega}}$
   &
 \\ \cline{2-3}
   & $\xi_0\gamma_{-1/2}b_{-1}\alpha_{-1}^\mu\ket{\wt{\Omega}}$
   &
 \\ \cline{2-3}
   & $\xi_0\psi_{-1/2}^\mu\alpha_{-1}^\nu\alpha_{-1}^\lambda
     \ket{\wt{\Omega}}$
   & $\xi c \psi^\mu e^{-\phi}:\p X^\nu\p X^\lambda e^{ikX}:$
 \\ \cline{2-3}
   & $\xi_0\psi_{-1/2}^\mu b_{-1}c_{-1}\ket{\wt{\Omega}}$
   & $\xi \p^2 c \psi^\mu e^{-\phi} e^{ikX}$
 \\ \hline
\end{tabular}
\caption{Basis for $\Phi$. Only those with $\xi_0$ and without $c_0$
are shown. For the level-2 basis, the operator expressions are given only for 
those relevant to our calculation.
}
\end{center}
\end{table}

\begin{table}[htbp]
\begin{center}
\begin{tabular}{|c|r|l|} \hline
 Level & Oscillator & Operator \\
       & expression & expression \\ \hline\hline
 0 & $\xi_0 c_0\beta_{-1/2}\ket{\wt{\Omega}}$ 
   & $\xi\p\xi c\p c e^{-2\phi}e^{2ik\cdot X}$
 \\ \hline\hline
 1 & $\xi_0 c_0\beta_{-3/2}\ket{\wt{\Omega}}$
   & $(\xi\p^2\xi c\p c e^{-2\phi}
     +\Half\xi\p\xi c\p c \p e^{-2\phi})e^{2ik\cdot X}$
 \\ \cline{2-3}
   & $\xi_0 c_0\beta_{-1/2}\beta_{-1/2}\gamma_{-1/2}\ket{\wt{\Omega}}$
   & $(\xi\p^2\xi c \p c e^{-2\phi}
     +\xi\p\xi c \p c \p e^{-2\phi})e^{2ik\cdot X}$
 \\ \cline{2-3}
   & $\xi_0 c_0\beta_{-1/2}\psi_{-1/2}^\mu\psi_{-1/2}^\nu\ket{\wt{\Omega}}$
   & $\xi\p\xi c\p c:\psi^\mu\psi^\nu:e^{-2\phi}e^{2ik\cdot X}$
 \\ \cline{2-3}
   & $\xi_0 c_0\beta_{-1/2}\alpha_{-1}^\mu\ket{\wt{\Omega}}$
   & $\xi\p\xi c\p c e^{-2\phi}:\p X^\mu e^{2ik\cdot X}:$
 \\ \cline{2-3}
   & $\xi_0 c_0\psi_{-1/2}^\mu b_{-1}\ket{\wt{\Omega}}$
   & $\xi \p c \psi^\mu e^{-\phi}e^{2ik\cdot X}$
 \\ \hline\hline
 2 & $\xi_0 c_0\beta_{-5/2}\ket{\wt{\Omega}}$
   & \\ \cline{2-3}
   & $\xi_0 c_0\beta_{-3/2}\psi_{-1/2}^\mu\psi_{-1/2}^\nu\ket{\wt{\Omega}}$
   & \\ \cline{2-3}
   & $\xi_0 c_0\beta_{-3/2}\beta_{-1/2}\gamma_{-1/2}\ket{\wt{\Omega}}$
   & \\ \cline{2-3}
   & $\xi_0 c_0\gamma_{-3/2}\beta_{-1/2}\beta_{-1/2}\ket{\wt{\Omega}}$
   & \\ \cline{2-3}
   & $\xi_0 c_0\psi_{-3/2}^\mu\beta_{-1/2}\psi_{-1/2}^\nu\ket{\wt{\Omega}}$
   & \\ \cline{2-3}
   & $\xi_0 c_0\beta_{-3/2}\alpha_{-1}^\mu\ket{\wt{\Omega}}$
   & \\ \cline{2-3}
   & $\xi_0 c_0\psi_{-3/2}^\mu b_{-1}\ket{\wt{\Omega}}$
   & \\ \cline{2-3}
   & $\xi_0 c_0\beta_{-1/2}\beta_{-1/2}\beta_{-1/2}\gamma_{-1/2}\gamma_{-1/2}
     \ket{\wt{\Omega}}$
   & \\ \cline{2-3}
   & $\xi_0 c_0\beta_{-1/2}\beta_{-1/2}\gamma_{-1/2}
     \psi_{-1/2}^\mu\psi_{-1/2}^\nu
     \ket{\wt{\Omega}}$
   & \\ \cline{2-3}
   & $\xi_0 c_0\beta_{-1/2}\psi_{-1/2}^\mu\psi_{-1/2}^\nu
     \psi_{-1/2}^\lambda\psi_{-1/2}^\rho\ket{\wt{\Omega}}$
   & \\ \cline{2-3}
   & $\xi_0 c_0\beta_{-1/2}\beta_{-1/2}\gamma_{-1/2}\alpha_{-1}^\mu
     \ket{\wt{\Omega}}$
   & \\ \cline{2-3}
   & $\xi_0 c_0\beta_{-1/2}\beta_{-1/2}\psi_{-1/2}^\mu c_{-1}
     \ket{\wt{\Omega}}$
   & \\ \cline{2-3}
   & $\xi_0 c_0\beta_{-1/2}\gamma_{-1/2}\psi_{-1/2}^\mu b_{-1}
     \ket{\wt{\Omega}}$
   & \\ \cline{2-3}
   & $\xi_0 c_0\beta_{-1/2}\psi_{-1/2}^\mu\psi_{-1/2}^\nu\alpha_{-1}^\lambda
     \ket{\wt{\Omega}}$
   & \\ \cline{2-3}
   & $\xi_0 c_0\psi_{-1/2}^\mu\psi_{-1/2}^\nu\psi_{-1/2}^\lambda b_{-1}
     \ket{\wt{\Omega}}$
   & \\ \cline{2-3}
   & $\xi_0 c_0\beta_{-1/2}\alpha_{-2}^\mu\ket{\wt{\Omega}}$
   & \\ \cline{2-3}
   & $\xi_0 c_0\psi_{-1/2}^\mu b_{-2}\ket{\wt{\Omega}}$
   & $\xi:bc\p c:\psi^\mu e^{-\phi} e^{ikX}$ \\ \cline{2-3}
   & $\xi_0 c_0\beta_{-1/2}\alpha_{-1}^\mu\alpha_{-1}^\nu\ket{\wt{\Omega}}$
   & \\ \cline{2-3}
   & $\xi_0 c_0\beta_{-1/2}b_{-1}c_{-1}\ket{\wt{\Omega}}$
   & \\ \cline{2-3}
   & $\xi_0 c_0\psi_{-1/2}^\mu b_{-1}\alpha_{-1}^\nu\ket{\wt{\Omega}}$
   & \\ \hline
\end{tabular}
\caption{Basis for $\Phi$. Only those with $\xi_0$ and $c_0$
are shown. For the level-2 basis, the operator expressions are given only for 
those relevant to our calculation.}
\end{center}
\end{table}

\begin{table}[htbp]
\begin{center}
\begin{tabular}{|c|r|l|} \hline
 Level & Oscillator & Operator \\
       & expression & expression \\ \hline\hline
 0 & $\xi_0\beta_{-1/2}\ket{\wt{\Omega}}$
   & $\xi\p\xi c e^{-2\phi} e^{ik\cdot X}$
 \\ \hline\hline
 1 & $\xi_0\beta_{-1/2}\psi_{-1/2}^\mu\psi_{-1/2}^\nu\ket{\wt{\Omega}}$
   & $\xi\p\xi c:\psi^\mu\psi^\nu:e^{-2\phi} e^{ik\cdot X}$
 \\ \cline{2-3}
   & $\xi_0\beta_{-1/2}\alpha_{-1}^\mu\ket{\wt{\Omega}}$
   & $\xi\p\xi c\p X^\mu e^{-2\phi} e^{ik\cdot X}$
 \\ \cline{2-3}
   & $\xi_0\psi_{-1/2}^\mu b_{-1}\ket{\wt{\Omega}}$
   & $\xi \psi^\mu e^{-\phi} e^{ik\cdot X}$
 \\ \cline{2-3}
   & $\xi_0\beta_{-3/2}\ket{\wt{\Omega}}$
   & $(\xi\p^2\xi c e^{-2\phi}
      +\Half\xi\p\xi c \p e^{-2\phi})e^{ik\cdot X}$
 \\ \cline{2-3}
   & $\xi_0\beta_{-1/2}\beta_{-1/2}\gamma_{-1/2}\ket{\wt{\Omega}}$
   & $(\xi\p\xi c \p e^{-2\phi}
      +\xi\p^2\xi c e^{-2\phi})e^{ik\cdot X}$
 \\ \hline\hline
 2 & $\xi_0\beta_{-5/2}\ket{\wt{\Omega}}$
   & $(\xi\p^3\xi e^{-2\phi}-2\xi\p^2\xi:\p\phi e^{-2\phi}:
      +\xi\p\xi:(\p\phi\p\phi-\p^2\phi)e^{-2\phi}:)c e^{ik\cdot X}$ 
 \\ \cline{2-3}
   & $\xi_0\beta_{-3/2}\beta_{-1/2}\gamma_{-1/2}\ket{\wt{\Omega}}$
   & $(\xi\p^2\xi \p e^{-2\phi}+\Half\xi\p^3\xi e^{-2\phi}
      +\Half\xi\p\xi:(\p\phi\p\phi-\p^2\phi)e^{-2\phi}:)
      ce^{ik\cdot X}$
 \\ \cline{2-3}
   & $\xi_0\beta_{-3/2}\alpha_{-1}^\mu\ket{\wt{\Omega}}$
   & $(\xi\p^2\xi e^{-2\phi}+\Half\xi\p\xi \p e^{-2\phi})
      c:\p X^\mu e^{ik\cdot X}:$
 \\ \cline{2-3}
   & $\xi_0\beta_{-3/2}\psi_{-1/2}^\mu\psi_{-1/2}^\nu\ket{\wt{\Omega}}$
   & $(\xi\p^2\xi e^{-2\phi}+\Half\xi\p\xi \p e^{-2\phi})
      c:\psi^\mu\psi^\nu:e^{ik\cdot X}$
 \\ \cline{2-3}
   & $\xi_0\beta_{-1/2}\beta_{-1/2}\beta_{-1/2}
      \gamma_{-1/2}\gamma_{-1/2}\ket{\wt{\Omega}}$
   & $(6\xi\p\xi:\p\phi\p\phi e^{-2\phi}:+3\xi\p^2\xi \p e^{-2\phi}
      +\xi\p^3\xi e^{-2\phi})ce^{ik\cdot X}$
 \\ \cline{2-3}
   & $\xi_0\beta_{-1/2}\beta_{-1/2}\gamma_{-3/2}\ket{\wt{\Omega}}$
   & $(\xi\p\xi:(\p\phi\p\phi+\p^2\phi)e^{-2\phi}:
      -\xi\p^2\xi:\p\phi e^{-2\phi}:)ce^{ik\cdot X}$
 \\ \cline{2-3}
   & $\xi_0\beta_{-1/2}\beta_{-1/2}\gamma_{-1/2}\alpha_{-1}^\mu
      \ket{\wt{\Omega}}$
   & $(\xi\p\xi \p e^{-2\phi}+\xi\p^2\xi e^{-2\phi})
      c:\p X^\mu e^{ik\cdot X}:$
 \\ \cline{2-3}
   & $\xi_0\beta_{-1/2}\beta_{-1/2}\gamma_{-1/2}
      \psi_{-1/2}^\mu\psi_{-1/2}^\nu\ket{\wt{\Omega}}$
   & $(\xi\p\xi \p e^{-2\phi}+\xi\p^2\xi e^{-2\phi})c
      :\psi^\mu\psi^\nu:e^{ik\cdot X}$
 \\ \cline{2-3}
   & $\xi_0\beta_{-1/2}\beta_{-1/2}c_{-1}\psi_{-1/2}^\mu\ket{\wt{\Omega}}$
   & $\xi\p\xi\p^2\xi c\p^2 ce^{-3\phi}\psi^\mu e^{ik\cdot X}$
 \\ \cline{2-3}
   & $\xi_0\beta_{-1/2}\alpha_{-2}^\mu\ket{\wt{\Omega}}$
   & $\xi\p\xi ce^{-2\phi}:\p^2 X^\mu e^{ik\cdot X}:$
 \\ \cline{2-3}
   & $\xi_0\beta_{-1/2}\alpha_{-1}^\mu\alpha_{-1}^\nu\ket{\wt{\Omega}}$
   & $\xi\p\xi ce^{-2\phi}:\p X^\mu\p X^\nu e^{ik\cdot X}:$
 \\ \cline{2-3}
   & $\xi_0\beta_{-1/2}\alpha_{-1}^\mu
      \psi_{-1/2}^\nu\psi_{-1/2}^\lambda\ket{\wt{\Omega}}$
   & $\xi\p\xi ce^{-2\phi}:\psi^\mu\psi^\nu:
      :\p X^\lambda e^{ik\cdot X}:$
 \\ \cline{2-3}
   & $\xi_0\beta_{-1/2}\psi_{-3/2}^\mu\psi_{-1/2}^\nu\ket{\wt{\Omega}}$
   & $\xi\p\xi ce^{-2\phi}:\p\psi^\mu \psi^\nu:e^{ik\cdot X}$
 \\ \cline{2-3}
   & $\xi_0\beta_{-1/2}\psi_{-1/2}^\mu\psi_{-1/2}^\nu
      \psi_{-1/2}^\lambda\psi_{-1/2}^\rho\ket{\wt{\Omega}}$
   & $\xi\p\xi ce^{-2\phi}:\psi^\mu\psi^\nu\psi^\lambda\psi^\rho:
      e^{ik\cdot X}$
 \\ \cline{2-3}
   & $\xi_0\beta_{-1/2}b_{-1}\gamma_{-1/2}\psi_{-1/2}^\mu\ket{\wt{\Omega}}$
   & $\xi \p e^{-\phi}\psi^\mu e^{ik\cdot X}$
 \\ \cline{2-3}
   & $\xi_0\beta_{-1/2}b_{-1}c_{-1}\ket{\wt{\Omega}}$
   & $\xi\p\xi \p^2 c e^{-2\phi} e^{ik\cdot X}$
 \\ \cline{2-3}
   & $\xi_0b_{-2}\psi_{-1/2}^\mu\ket{\wt{\Omega}}$
   & $\xi:bc:e^{-\phi}\psi^\mu e^{ik\cdot X}$
 \\ \cline{2-3}
   & $\xi_0b_{-1}\psi_{-3/2}^\mu\ket{\wt{\Omega}}$
   & $\xi e^{-\phi}\p\psi^\mu e^{ik\cdot X}$
 \\ \cline{2-3}
   & $\xi_0b_{-1}\psi_{-1/2}^\mu\alpha_{-1}^\nu\ket{\wt{\Omega}}$
   & $\xi e^{-\phi}\psi^\mu:\p X^\nu e^{ik\cdot X}:$
 \\ \cline{2-3}
   & $\xi_0b_{-1}\psi_{-1/2}^\mu\psi_{-1/2}^\nu\psi_{-1/2}^\lambda
      \ket{\wt{\Omega}}$
   & $\xi e^{-\phi}:\psi^\mu\psi^\nu\psi^\lambda:e^{ik\cdot X}$
 \\ \cline{2-3}
   & $b_{-2}b_{-1}\ket{\Omega}$ 
   & $be^{ik\cdot X}$
 \\ \hline
\end{tabular}
\caption{Basis for $\Omega$. Only those without $c_0$ are shown.}
\end{center}
\end{table}

\begin{table}[htbp]
\begin{center}
\begin{tabular}{|c|r|l|} \hline
 Level & Oscillator & Operator \\
       & expression & expression \\ \hline\hline
 0 & \mbox{none} & \\ \hline\hline
 1 & $\xi_0 c_0\beta_{-1/2} b_{-1}\ket{\wt{\Omega}}$
   & $\xi\p\xi \p c e^{-2\phi} e^{ik\cdot X}$
 \\ \cline{2-3}
   & $\xi_0 c_0\beta_{-1/2}\beta_{-1/2}\psi_{-1/2}^\mu\ket{\wt{\Omega}}$
   & $\xi\p\xi\p^2\xi c \p c \psi^\mu e^{-3\phi} e^{ik\cdot X}$
 \\ \hline\hline
 2 & $\xi_0 c_0\beta_{-3/2}\beta_{-1/2}\psi_{-1/2}^\mu\ket{\wt{\Omega}}$
   & $(\xi\p\xi\p^3\xi e^{-3\phi}-2\xi\p\xi\p^2\xi:\p\phi e^{-3\phi}:)
      c\p c \psi^\mu e^{ik\cdot X}$
 \\ \cline{2-3}
   & $\xi_0 c_0\beta_{-3/2}b_{-1}\ket{\wt{\Omega}}$
   & $(\xi\p^2\xi e^{-2\phi}+\Half\xi\p\xi \p e^{-2\phi})
      \p c e^{ik\cdot X}$
 \\ \cline{2-3}
   & $\xi_0 c_0\beta_{-1/2}\beta_{-1/2}\beta_{-1/2}
      \gamma_{-1/2}\psi_{-1/2}^\mu\ket{\wt{\Omega}}$
   & $(\xi\p\xi\p^2\xi \p e^{-3\phi}+\xi\p\xi\p^3\xi e^{-3\phi})
      c \p c \psi^\mu e^{ik\cdot X}$
 \\ \cline{2-3}
   & $\xi_0 c_0\beta_{-1/2}\beta_{-1/2}\beta_{-1/2}c_{-1}\ket{\wt{\Omega}}$
   & $\xi\p\xi\p^2\xi\p^3\xi e^{-4\phi}
      c \p c \p^2 c e^{ik\cdot X}$
 \\ \cline{2-3}
   & $\xi_0 c_0\beta_{-1/2}\beta_{-1/2}\psi_{-3/2}^\mu\ket{\wt{\Omega}}$
   & $\xi\p\xi\p^2\xi e^{-3\phi}
      c \p c \p\psi^\mu e^{ik\cdot X}$
 \\ \cline{2-3}
   & $\xi_0 c_0\beta_{-1/2}\beta_{-1/2}
      \psi_{-1/2}^\mu\psi_{-1/2}^\nu\psi_{-1/2}^\lambda\ket{\wt{\Omega}}$
   & $\xi\p\xi\p^2\xi e^{-3\phi}
      c \p c :\psi^\mu\psi^\nu\psi^\lambda:e^{ik\cdot X}$
 \\ \cline{2-3}
   & $\xi_0 c_0\beta_{-1/2}\beta_{-1/2}
      \psi_{-1/2}^\mu\alpha_{-1}^\nu\ket{\wt{\Omega}}$
   & $\xi\p\xi\p^2\xi e^{-3\phi}
      c \p c \psi^\mu:\p X^\nu e^{ik\cdot X}:$
 \\ \cline{2-3}
   & $\xi_0 c_0\beta_{-1/2}\beta_{-1/2}b_{-1}\gamma_{-1/2}\ket{\wt{\Omega}}$
   & $(\xi\p\xi \p e^{-2\phi}+\xi\p^2\xi e^{-2\phi})
      \p c e^{ik\cdot X}$
 \\ \cline{2-3}
   & $\xi_0 c_0\beta_{-1/2}b_{-2}\ket{\wt{\Omega}}$
   & $\xi\p\xi e^{-2\phi}:bc \p c:e^{ik\cdot X}$
 \\ \cline{2-3}
   & $\xi_0 c_0\beta_{-1/2}b_{-1}\alpha_{-1}^\mu\ket{\wt{\Omega}}$
   & $\xi\p\xi e^{-2\phi} \p c :\p X^\mu e^{ik\cdot X}:$
 \\ \cline{2-3}
   & $\xi_0 c_0\beta_{-1/2}b_{-1}\psi_{-1/2}^\mu\psi_{-1/2}^\nu
      \ket{\wt{\Omega}}$
   & $\xi\p\xi e^{-2\phi} \p c:\psi^\mu\psi^\nu:e^{ik\cdot X}$
 \\ \hline
\end{tabular}
\caption{Basis for $\Omega$. Only those with $c_0$ are shown.}
\end{center}
\end{table}

}

\newcommand{\J}[4]{{\sl #1} {\bf #2} (#3) #4}
\newcommand{\andJ}[3]{{\bf #1} (#2) #3}
\newcommand{\AP}{Ann.\ Phys.\ (N.Y.)}
\newcommand{\MPL}{Mod.\ Phys.\ Lett.}
\newcommand{\NP}{Nucl.\ Phys.}
\newcommand{\PL}{Phys.\ Lett.}
\newcommand{\PR}{Phys.\ Rev.}
\newcommand{\PRL}{Phys.\ Rev.\ Lett.}
\newcommand{\PTP}{Prog.\ Theor.\ Phys.}
\newcommand{\hepth}[1]{{\tt hep-th/#1}}

\end{document}